\documentclass[]{emulateapj}

\def\cm3{\mbox{cm$^{-3}$}}
\def\erg{\mbox{ergs}}
\def\Ke{\mbox{K}}
\def\kpc{\mbox{kpc}}
\def\kpch{\mbox{$h^{-1}$kpc}}
\def\LCDM{{$\Lambda$CDM}}
\def\mpch{\mbox{$h^{-1}$Mpc}}

\def\msun{\mbox{M$_\odot$}}
\def\msunpc{\mbox{M$_\odot$ pc$^{-2}$}}
\def\msunh{\mbox{$h^{-1}$M$_\odot$}}    
\def\mv{\mbox{$M_{\rm vir}$}}

\def\ome{\mbox{$\Omega_0$}}
\def\omel{\mbox{$\Omega_\Lambda$}}
\def\omeb{\mbox{$\Omega_b$}}

\def\rv{\mbox{$R_{\rm vir}$}}
\def\re{\mbox{$R_{e}$}}

\def\eg{e.g.,}
\def\nbody{\mbox{$N-$body}}
\def\yr{\mbox{yr}}

\def\sigSFR{\mbox{$\Sigma_{\rm SFR}$}}
\def\siggas{\mbox{$\Sigma_{g}$}}

\def\ltsima{$\; \buildrel < \over \sim \;$}    
\def\lesssim{\lower.5ex\hbox{\ltsima}}           
\def\gtsima{$\; \buildrel > \over \sim \;$}    
\def\gtrsim{\lower.5ex\hbox{\gtsima}}           

\newcommand{\Dvir}{\Delta_{\rm vir}}
\newcommand{\kms}{{\rm ~km~s}^{-1}}
\newcommand{\lmax}{l_{\rm max}}
\newcommand{\mlim}{m_{{\rm *,lim}}}
\newcommand{\mst}{m_{{\rm *}}}

\newcommand{\nsf}{n_{\rm SF}}
\newcommand{\nH}{n_{\rm H}}
\newcommand{\pcc}{{\rm cm}^{-3}}
\newcommand{\psc}{{\rm cm}^{-2}}
\newcommand{\rhog}{\rho_{\rm g}}
\newcommand{\rhosf}{\rho_{\rm SF}}
\newcommand{\Tsf}{T_{\rm SF}}
\newcommand{\Vtot}{V_{\rm tot}}
\newcommand{\VS}{V{\'a}zquez-Semadeni}
\newcommand{\BP}{Ballesteros-Paredes}

\def\spose#1{\hbox to 0pt{#1\hss}}
\newcommand\lsim{\mathrel{\spose{\lower 3pt\hbox{$\mathchar"218$}}
     \raise 2.0pt\hbox{$\mathchar"13C$}}}
\newcommand\gsim{\mathrel{\spose{\lower 3pt\hbox{$\mathchar"218$}}
     \raise 2.0pt\hbox{$\mathchar"13E$}}}

\begin{document}

\title{Low--mass galaxy formation in cosmological AMR simulations: 
the effects of varying the sub--grid physics parameters}

\author{Pedro Col\'{\i}n$^{1}$
Vladimir Avila-Reese$^2$, Enrique V\'azquez-Semadeni$^1$, Octavio Valenzuela$^2$,
and Daniel Ceverino$^3$}
\affil{$^1$ Centro de Radioastronom\'{\i}a y Astrof\'{\i}sica, Universidad Nacional 
Aut\'onoma de M\'exico, A.P. 72-3 (Xangari), Morelia, Michoac\'an 58089, M\'exico
\\ $^2$ Instituto de Astronom\'{\i}a, Universidad Nacional Aut\'onoma de M\'exico, 
A.P. 70-264, 04510, M\'exico, D.F., M\'exico \\
$^3$ Racah Institute of Physics, The Hebrew University, Jerusalem 91904, Israel}


\email{p.colin@crya.unam.mx}

\begin{abstract} 
We present numerical simulations aimed at exploring the effects of varying the 
sub--grid physics parameters on the evolution and the properties of the galaxy 
formed in a low--mass dark matter halo ($\sim 7 \times 10^{10}$ \msunh\ at redshift $z=0$). 
The simulations are run within a cosmological setting with a nominal resolution of
218 pc comoving and are stopped at $z = 0.43$. For simulations that
cannot resolve individual molecular clouds, we propose the criterion 
that the threshold density for star formation, $\nsf$, should be chosen
such that the column density of the star-forming cells 
equals the threshold value for molecule formation, $N \sim 10^{21}$ cm$^{-2}$, or 
$\sim 8\ \msunpc$. 
In all of our simulations, an extended 
old/intermediate--age stellar halo 
and a more compact younger stellar disk are formed,
and in most cases, the halo's
specific angular momentum is slightly larger than that of the galaxy, and
sensitive to the SF/feedback parameters. We found that a non negligible 
fraction of the halo stars are formed in situ in a spheroidal distribution. 
Changes in  
the sub--grid physics parameters affect significantly and in a complex way 
the evolution and properties of the galaxy: 
(i) Lower threshold densities $\nsf$ produce larger stellar effective radii \re, 
less peaked circular velocity curves $V_c(R)$, and greater amounts of 
low-density and hot gas in the disk mid-plane; 
(ii) When stellar feedback is modeled 
by temporarily switching off
radiative cooling in the star forming regions, \re\ increases (by 
a factor of $\sim 2$ in our particular model), 
 the circular velocity curve becomes flatter, 
and a complex multi--phase gaseous disk structure develops;
(iii) A more efficient local conversion of gas mass to stars, measured by a
stellar particle mass distribution biased toward larger values,
increases the strength of the feedback  energy injection  --driving 
outflows and inducing  burstier SF histories;
iv) If feedback is too strong, gas loss by galactic outflows ---which
are easier to produce in low--mass galaxies---  
interrupts SF, whose history becomes episodic;
v) In all cases, the
surface SF rate versus the gas surface density correlation  
is steeper than the Kennicutt law but in agreement with observations in low-surface 
brightness galaxies. The simulations exhibit two important shortcomings: 
the baryon fractions are higher, and the specific SF rates are much smaller, than
observationally inferred values for redshifts $\sim 0.4-1$. 
These shortcomings pose a major challenge to the SF/feedback physics commonly applied
in the \LCDM--based galaxy formation simulations.
\end{abstract}

\section{Introduction}

The succesful development of cosmology in the last two decades,
incarnated in the popular \LCDM\ model, has provided a robust
theoretical background for modeling galaxy formation and evolution
\citep[for reviews see][]{Baugh06,A-R06}. 
Due to the high non-linearity implied in the problem, cosmological 
N-body + hydrodynamical simulations offer the fairest way to attain
such a modeling \citep[for a recent review see][]{Mayer08}. 
However, this method is hampered by the large --currently unreachable-- 
dynamic range required to model galaxy formation and evolution in 
the cosmological context (formally from the subpc scales of molecular cloud cores 
to Mpc scales, and from densities $n \sim 10^{-6}$ to $\gtrsim 10^{2}\ \pcc$), 
as well as by the complexity of the processes involved, mainly 
cooling and star 
formation (SF) and its feedback on the surrounding medium. Both of them occur 
at scales commonly well below the accesible resolution in simulations.  
Theoretical and observational results suggest that the star formation efficiency can be simulated
without reaching the subparsec resolution if  a simulation can resolve the mean 
density of giant molecular clouds \citep{Gnedin09, Krumholz05}, unfortunately 
this seems not to be the case 
for SN and Stellar Wind feedback.
Undoubtedly, modeling of SF and feedback (subgrid physics) is of crucial 
importance for the simulations of galaxy formation and evolution.

On the other hand, the SF rate (SFR) and its evolution are among the most important
observational properties of galaxies, being basically  responsible for coining 
the present-day galactic stellar populations.
Star formation is also crucial in terms of its feedback effect, which (i) 
regulates several thermo- and hydro-dynamical processes of the multi-phase 
ISM, including probably the SF itself; (ii) partially controls the 
galaxy assembly process through mass outflows and re-heating of the circumgalactic 
medium, and (iii) likely is also the driving agent of the intergalactic 
chemical enrichment.  

In the context of the hierarchical \LCDM\ scenario, 
the efficient radiative cooling of gas leads to the so-called 
``gas cooling catastrophe'': 
most of the available primordial gas is cooled, trapped, and 
transformed into stars inside the smallest, early-collapsing halos 
\citep{whiterees78}. In order to avoid such a problem, the stellar negative 
feedback is routinely invoked \citep[see for recent works e.g.,][ though alternative
physical processes have been also considered, e.g., Sommer-Larsen \& Dolgov 2001;
Governato et al. 2004]{Okamoto05, Zavala08}. 
In the same vein, it is well known now that the mass fraction of 
baryons in galaxies with respect to their halo masses is much smaller than the 
universal baryon fraction $f_b = \omeb/\ome$.
The `missing baryons' are commonly assumed to have been expelled from galaxies
by feedback effects, especially in the past, when the halos were 
smaller \citep{DekelSilk86}.

In the absence of a general theory of SF and feedback, several 
prescriptions have been proposed in order to model the corresponding subgrid 
physics in numerical simulations  
at parsec scales or smaller, with the aim to properly extend 
SF/feedback influence into the larger (tens/hundreds of parsecs)    
resolved scales. Unavoidably, subgrid SF and feedback prescriptions in 
simulations introduce several free parameters. 

In simulations, some of the common criteria used for declaring 
a gas cell or particle 
as star forming are converging cold gas flows, a local Jeans instability 
and/or a  
local gas density threshold ($\nsf$) \citep[see e.g.,][]{KWH96,Kravtsov03}.  
The rate at which the gaseous mass element obeying the above criteria is 
converted into `stars' is calculated using recipes which introduce some 
free parameters like the SF efficieny constant $C_*$ 
\citep[e.g.,][]{Katz92,NavarroWhite93,GI97,SpringelHernquist03}.  
In order to take into account the stochastic nature of gas conversion into 
stars, and also the fact that the minimum resolution element overlooks local  
variations in gas density, a probabilistic function around $\nsf\ $ is 
typically introduced instead of a deterministic density criterion
\citep[e.g.,][]{Stinson06}.

For the modeling of the feedback processes associated to the non-resolved 
regions, in which they start to operate, two schemes are 
commonly used: individual gas elements surrounding a stellar particle are 
given a velocity kick \citep[kinematic feedback; e.g.,][]{NavarroWhite93,Abadi03,DT2008}
and/or are injected with thermal energy and metals \citep[e.g.,][]{Kravtsov03},
at the same time that, in some schemes, the radiative cooling is 
temporarly turn off to prevent the thermal energy from being radiated away
before it is converted into kinetic energy \citep[adiabatic feedback; e.g.,][]{GI97,
TC00,Stinson06}. Alternatively, in other schemes, the interaction between
neighbor particles belonging to different phases is inhibited by hand 
\citep{Okamoto03,Scannapieco06}. 
A third scheme, often applied, imposes an effective gas equation of state
in order to describe the unresolved multiphase ISM \citep[e.g.,][]{Yepes97,
SpringelHernquist03}. Hybrid schemes were also used: semi-analytical models 
of winds and starburts dependent on the gas physical conditions are included 
in the simulations \citep{SpringelHernquist03}.
In all these feedback schemes there are free parameters, such as the amounts 
of momentum and thermal energy deposited into the gas elements, the time the 
cooling is switched off, $t_{\rm off}$, etc. 

The values for the free parameters in the SF and feedback prescriptions are chosen 
in different ways. They can be defined {\it a priori}, based on constraints typically 
related to observations of the Galaxy ISM, and/or {\it a posteriori}, choosing 
the parameter 
values that make the evolution and properties of the simulated galaxy close to the 
observed ones. The prescriptions and parameters used also depend on whether the
code is of the smoothed particle hydrodynamics (SPH) or Eulerian 
adaptive mesh refinement (AMR) kind.

Because of the large diversity of models, strategies, and relevant parameters 
of SF/feedback implementations, a comparison between them is not trivial. 
Although some implementations in high--resolution simulations have reached 
an amount of success forming rotational supported disk galaxies  
\citep[see, for recent reviews,][ and references therein]{Mayer08,Gibson09}, 
the task is far from being completed. 
It is not clear what are the quantitative differences between each subgrid model. 
Thus, a characterization of the various subgrid schemes for the 
ISM energy injection and dissipation is needed to gain some insight 
into their ability/inability to form disk galaxies.
 
In this paper, we explore the effects of the SF/feedback parameters on 
the structure and evolution of a 
{\it low-mass galaxy} 
in cosmological simulations using the Hydrodynamics Adaptive Refinement Tree (ART) 
code \citep{KKK97,Kravtsov03}.  
We reach a maximum resolution of about one hundred parsecs at redshift $z=1$ in order to 
be able to resolve the clouds associated with the SF; most of our analysis 
is made at this epoch and at $z=0.43$, the last epoch reached by all the runs.  
Our nominal resolution extrapolated to $z=0$ is of 218 pc, which is 
among the highest resolutions for a cosmological galaxy formation experiment
using the AMR/Eulerian technique \citep[c.f.][]{Gibson09}.

We will focus our analysis on general evolutionary and physical properties/processes 
of the simulated galaxies, namely: (i) the evolution of structural and dynamical 
properties (e.g., galaxy stellar structure, circular velocity curve decomposition), 
(ii) the structure/thermodynamics of the ISM (density and temperature probability 
distribution functions, vertical filling factors of the different phases of the ISM,
gas fractions), (iii) the dependence of the global SF history and the SFR
on gas density,
and (iv) global quantities like the baryon mass and angular momentum fractions. 
Our aim here is not to simulate galaxies that can be compared to observations,
but rather to determine the effects of the parameters of the SF/Feedback 
prescriptions. Moreover,
we will use our results to point out  some potential difficulties of simulated
low--mass galaxies in the context of the \LCDM\ cosmogony.     

The plan of the paper is as follows. The code and the SF and feedback prescriptions used for 
the simulations are described in \S 2. In this section, the cosmological simulation and the 
different runs varying the SF/feedback parameters are also presented. The results of
the various runs are given and compared with each other in \S 3. In \S 4 
we present a summary of the main results obtained in the paper accompanied by 
interpretations and discussions. Some potential vexing problems of any \LCDM-based model
of low--mass galaxy formation and evolution are hihglighted in \S 4.3. Finally,
our conclusions are given in  \S 5.

\section{Numerical Experiments}

\subsection{The Code} \label{sec:the_code}

The numerical simulations described here were performed using the hydrodynamics +
N-body Adaptive Refinement Tree code ART \citep[]{KKK97,Kravtsov03}. 
Among the physical 
processes included in the code are the cooling of the gas and its 
subsequent conversion
into stars, stellar feedback, self-consistent advection of metals,
an UV heating background source, etc.

The cooling and heating rates incorporate Compton heating/cooling, atomic and molecular 
cooling, UV heating from a cosmological background radiation \citep{HM96}, and are
tabulated for a temperature range of $10^2 < T < 10^9\ \Ke$ and a grid of densities,
metallicities, and redshifts using the CLOUDY code \citep[ version 96b4]{Ferland98}. 
Gas is converted into stellar particles according to a prescription described and 
discussed below.
Stellar particles eject metals and thermal energy through stellar winds and 
type II and Ia supernovae explosions. Each supernova dumps $2 \times 10^{51}\ \erg$
of thermal energy and ejects $1.3 \msun$ of metals. For the assumed \citet{MS79} initial
mass function, IMF, a stellar particle of $10^5\ \msun$ produces 749 type II supernovae.
For a more detailed discussion of the processes implemented in the code, see 
\citet[][]{Kravtsov03, KNV05}.

\subsection{Star formation and feedback} \label{sec:SF_n_feedback}

In ART, SF is modeled as taking place in the
coldest and densest collapsed regions, defined by $T < \Tsf$ and $\rho_g >
\rhosf$, where $T$ and $\rho_g$ are the temperature and density of
the gas, respectively, and $\rhosf$ is a density threshold. A stellar
particle of mass $m_*$ is placed in a grid cell where these conditions
are simultaneously satisfied, and this mass is removed from the gas mass
in the cell. The particle subsequently follows N-body dynamics. No other
criteria are imposed. 

Although $\nsf$, the hydrogen number density threshold corresponding to
$\rhosf$, is in principle a free parameter of the simulations, one can
make an ``educated guess'' as to which values should be considered most
realistic, by noting that SF occurs almost exclusively in giant
molecular clouds (GMCs). Thus, it is natural to require that the
density in a star-forming cell should be at least such that the cell's
column density equals the threshold value for molecular Hydrogen
and CO formation, $N \gtrsim 10^{21} \psc$ \citep{FC86, vDB88}. For a cell of 150 pc, this
translates into a number density $\nsf \sim 5 \pcc$. In our models,
$\nsf$ is varied around this value from 0.1 to 50 cm$^{-3}$. We keep
$\Tsf$ fixed at 9000 K.

The stellar particle mass, $m_*$, is determined through a simple
two-step process. A first estimate of this mass, labeled $m_1$, is made
by assuming that $m_*$ is proportional to a power law of the gas
density, $\rho_g^\alpha$. 
In this work we use $\alpha = 1.0$. Thus, in this step, collisionless stellar particles can
be created with a mass $m_1 = C_*~m_{\rm gas}~dt_0$, where $C_*$ is a
constant that measures the ``efficiency'' with which gas is locally
converted into stars during $dt_0$, the time step of the coarsest grid, 
which is also the time interval between successive checks by the code for the formation
of new stellar particles; $m_{\rm gas}$ is the gas mass in the cell. For low 
values of $C_*$, for example $C_* = 2.5 \times 10^{-10}\ \yr^{-1}$ (the fiducial
value in the code), the formula above produces small $m_1$
values. However, this makes the number of stellar particles in the
simulation practically intractable. For this reason, a stellar particle
mass limit is introduced in the SF set of parameters, labeled
$\mlim$. The SF algorithm then
chooses the maximum between $m_1$ and $\mlim$, which we call
$m_2$. In the second step, the algorithm takes the minimum between
$m_2$ and 2/3 of the gas mass contained in the cell and assigns this
mass to the stellar particle, in order to prevent total exhaustion of
the gas contents of the cell. Since the stellar particle masses are
 much more massive than the mass of a star, typically 
$10^4$ -- $10^5~\msun$, 
once formed, each stellar particle is
considered as a single stellar population, within which the individual
stellar masses are distributed according to the Miller \& Scalo IMF.

The simulations reported here are of two kinds: those with a
``deterministic'' SF prescription (the ``D'' runs), in
which stellar particles are always created once the density 
and temperature conditions are
satisfied, and those with a ``stochastic'' or ``random'' SF
prescription (the ``R'' runs ), in which stellar particles are created
in a cell with a probability given by
\begin{equation}
P = \left\{\begin{array}{ll}
          0  & \mbox{if}~\rhog < \rhosf;\\
          \frac{\rhog}{\beta \rhosf} & \mbox{if}~\rhosf < \rhog < \beta
               \rhosf;\\
          1  & \mbox{if}~\beta \rhosf < \rhog.
          \end{array} \right.
\label{eq:prob_SF}
\end{equation}
where $\beta$ is a free parameter, to which we refer as
the probability-saturation threshold.
Thus, regions with higher densities have higher probabilities to host
SF events. This stochastic prescription allows for the
possibility of forming stars in regions of low average density (assuming we have
chosen a low $\nsf$ parameter), in which isolated dense clouds would not
be resolved.

The runs can also be divided in two classes according to the stellar
feedback implementation. In the first class of runs, all the energy
from stellar feedback is dumped onto the star-forming cell
in the form of heat, as
described, for example, in \citet{Kravtsov03}. This thermal
feedback suffers from the well-known overcooling problem in low-resolution
simulations. This problem can be traced back to insufficient numerical
resolution \citep{CK2009}, and happens when the characteristic
expansion time in a cell heated by a stellar particle (given by the
sound crossing time across the cell) is much longer than the
cooling time at high densities and low resolutions (large cell
size). In this case, in a single timestep, the gas cools before it has
time to expand. The resolution condition to avoid this problem
would then be to require that the sound crossing time across the cell
be shorter than the local cooling time. This, however, can require
impossibly high resolutions with presently available computational
resources. 

A frequent strategy to overcome this problem \citep[e.g., ][]{GI97,
TC00, Sommer-Larsen03, Keres05, Governato07}
is to turn off the cooling in a cell for some time after a stellar
particle is produced there, in order to allow for the gas to expand
before it cools. Although this procedure has been criticized as being
somewhat arbitrary and unrealistic \citep[e.g., ][]{CK2009}, we 
consider that it is physically motivated,
as it is even more
unrealistic to prevent the expansion of heated regions due to the
inability to resolve the scales that would allow the physically correct
behavior of these regions. 
In the present paper we therefore adopt this
strategy, and test its effects on the evolution of the forming
galaxy. Specifically, we turn off the cooling in a cell for 
40 Myr after a SF event occurs there in some of the runs
(see Table \ref{tab:Models}).

\subsection{The numerical method}

We present here several simulations of the evolution of a
galaxy with a total (baryon + dark matter) mass of about $7 \times 10^{10}\ \msunh$ at the
present day. The analysis is made at two epochs, $z = 1$ and $z = 0.43$.
The latter corresponds to a time just before the last
major merger. The simulations are run in a \LCDM\ 
cosmolgy with $\ome = 0.3$, $\omel = 0.7$, and $\omeb = 0.045$. The CDM power spectrum
is taken from \citet{kh97} and it is normalized to $\sigma_8 = 0.8$,
where $\sigma_8$ is the rms amplitude of mass fluctuations in 8 \mpch\
spheres.

To maximize the resolution efficiency, we first perform a low-resolution
N-body simulation with $128^3$ dark matter (DM) particles in a periodic
box of $10 \mpch$ on a side. At the start of the simulation, the box is
initially covered by a mesh of $128^3$ cells (zeroth level cells). A
spherical region, centered on a {\it randomly} selected halo of mass
$2.6 \times 10^{11} \msunh$ and with a radius equal to three times the
virial radius (\rv) of the halo, was then chosen at $z = 0$ and its
corresponding Lagrangian region, identified at $z = 50$, was re-sampled
with additional small-scale modes \citep{KKBP01}. The virial radius is
defined as the radius that encloses a mean density equal to
$\Dvir$ times the mean density of the universe, where $\Dvir$
is a quantity that depends on  $\ome$, $\omel$ and $z$. For
example, for our cosmology $\Dvir(z=1) = 203$.  The number of DM
particles in the high-resolution zone is about one million and the mass
per particle ($m_{dm}$) is $5.3 \times 10^5 \msunh$.

In ART, the initially uniform grid is refined recursively as the matter
distribution evolves. The criterion chosen for refinement is based on gas or
dark matter densities. The cell is refined
when the mass in DM particles exceeds 1.3$(1 - f_b) m_p$ or
the mass in gas is higher than 13.0$f_b m_p$, where $m_p = m_{dm}/(1-f_b)$.  
For the
simulations presented in this paper, using multiple dark matter particle
masses, the grid is always unconditionally
refined to the third level, corresponding to an effective grid size of
$512^3$.  On the other hand, the maximum allowed refinement level
$\lmax$ was set to 9. At $z = 1$, the number of grid cells, at all
levels of resolution, is about 11 million.  In particular, at
$\lmax$ the number of cells is $\sim\ 5 \times 10^5$, corresponding to a
resolution of 218 comoving parsecs.

\begin{deluxetable*}{cccccc}
\tablecolumns{6}
\tablewidth{0pc}
\tablecaption{Parameter simulations}
\tablehead{\colhead{Model} & \colhead{$\mlim$\tablenotemark{a}} & \colhead{$\nsf$
\tablenotemark{b}} & \colhead{$C_*$\tablenotemark{c}} & 
\colhead{Cooling\tablenotemark{d}} & \colhead{$\beta$\tablenotemark{e}}  \\
  & ($10^4\ \msun$) & (cm$^{-3}$) &  ($2.5 \times 10^{-10}$ yr$^{-1}$) &  &
($10^6$ yr) }
\startdata
  D1     & 1.0      &  50.0  &    1.0 & on    & off    \\
  D2     & 8.0      &   6.25 &    1.0 & on    & off    \\
  D3     & 8.0      &   6.25 &    1.0 & 40    & off    \\
  D4     & $\infty$ &   6.25 &   none & 40    & off    \\
  R1     & 1.0      &   1.0  &  667.0 & on    & 100.0  \\
  R2     & 1.0      &   0.1  &  500.0 & 40    & 20.0   \\
  R3     & 1.0      &   1.0  &  500.0 & 40    & 20.0   \\
  R4     & 1.0      &   6.25 &  500.0 & 40    & 20.0   \\
  Ad     & off      &   off  &   off  & off   & off    \\	
\enddata
\tablecomments{Runs are denoted with the capital letters D or R
followed by a number. D stands for
deterministic and R for random. The adiabatic run is 
denoted by Ad. This simulation does not include radiative cooling.}

\tablenotetext{a}{Stellar particle mass limit.}
\tablenotetext{b}{Threshold gas density for star formation.}
\tablenotetext{c}{Star formation efficieny constant.}
\tablenotetext{d}{Cooling is always on or turned off for 40 Myr after SF events.}
\tablenotetext{e}{Probability-saturation threshold.}
\label{tab:Models}
\end{deluxetable*}

In this paper, we focus our study on the third most massive halo
appearing in the re-sampled simulations at $z = 1$ because it turns out
that it evolves from $z \sim 2$ to $z \sim 0.4$ in an nearly isolated fashion. 
This allow us to observe the effects that the SF and feedback
prescriptions have on the structure and evolution of the embedded galaxy
without having to deal with complicated dynamical interactions. This
halo contains within its virial radius $\sim 10^5$ DM particles at $z =
0$. In the remainder of the paper, we study the properties of this system
as a function of the SF/feedback prescriptions
and the corresponding parameter values. 

\subsection{The runs}

\begin{figure}[htb!]
\plotone{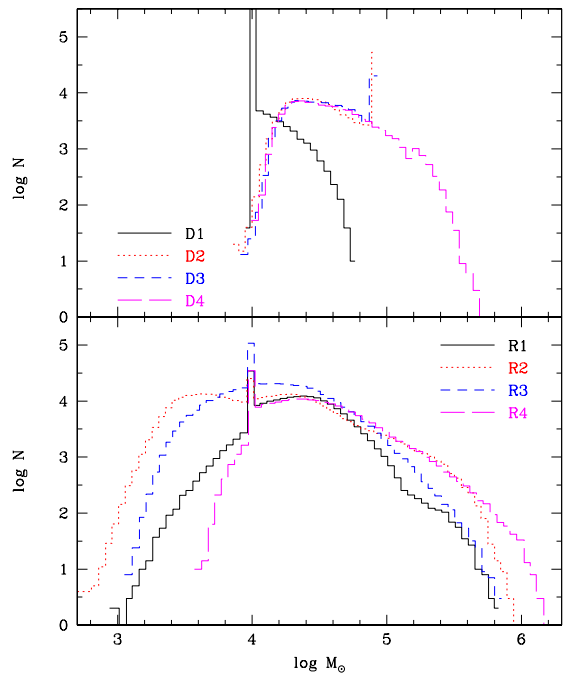}
\caption[fig:MstarDist]{Stellar particle mass distributions measured at $z=1.0 $
for all simulations presented in Table ~\ref{tab:Models}. The distribution
depends on the values of the SF parameters $\mlim$, $C_*$, and $\nsf$. 
\label{fig:MstarDist}}
\end{figure}

\begin{figure*}
\includegraphics[width=\textwidth]{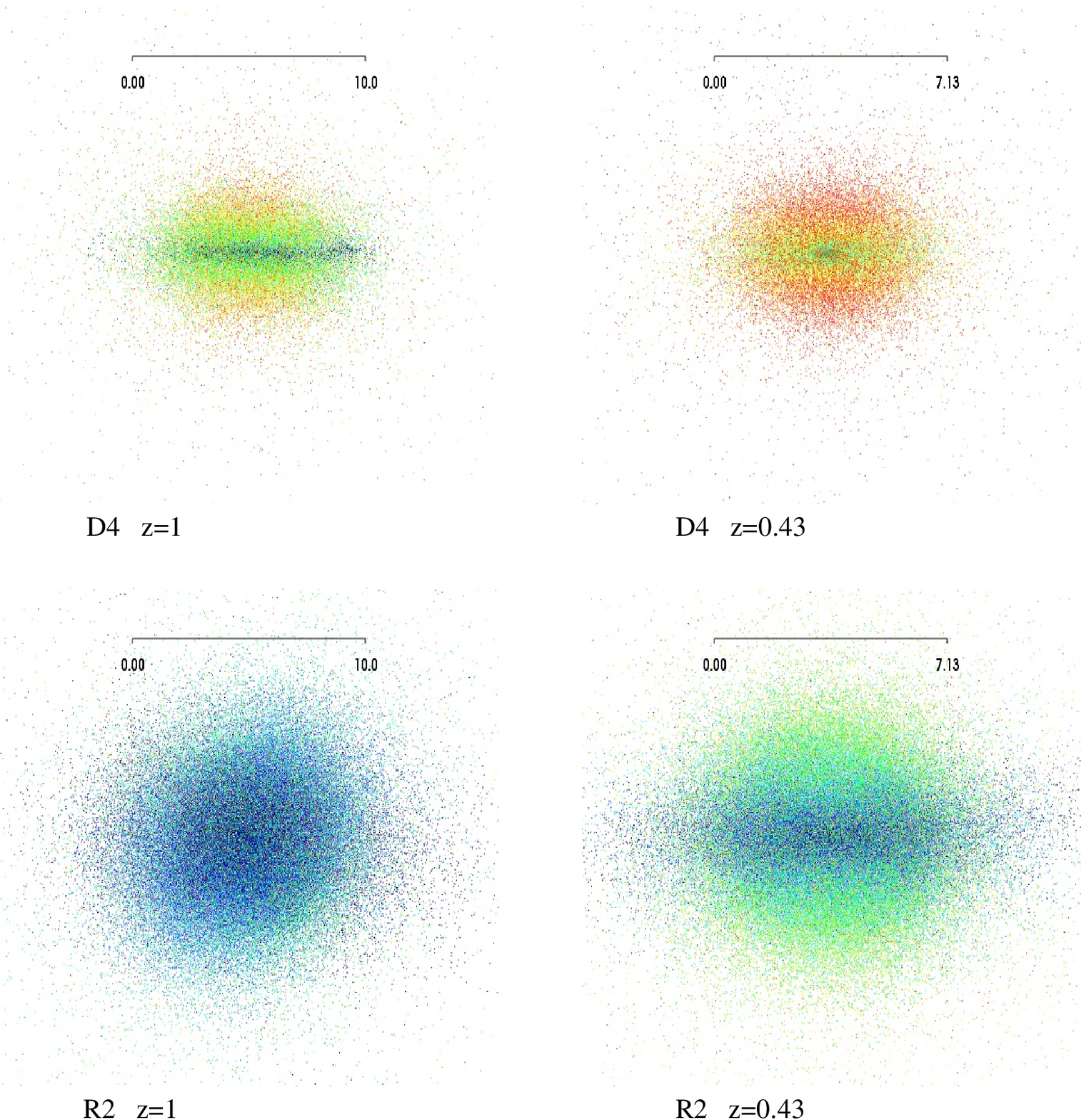} 
\caption[fig:Diskz=1]{Spatial distribution of stellar particles for
runs D4 and R2 at $z=1$ and $z=0.43$; the disks, when present, are 
shown edge on. 
Scale is comoving \kpch\ (solid line measures 10 \kpch). Particles are 
color coded by age with a rainbow palette
with the bluest particles being the youngest ones. At $z = 1$ in 
run D4 a young stellar disk can be clearly distinguished while in run R2 
the young stellar particles are distributed in a spheroid.
\label{fig:Diskz=1}}
\end{figure*}

In Table~\ref{tab:Models} we list the parameters described above for the
various runs we performed. The first column gives the name of the run, denoted by
a capital letter followed by a number, where the former indicates
whether the SF model belongs to the D or R sequence (see \S
\ref{sec:SF_n_feedback}). In columns (2), (3), and (4) we show the
values of $\mlim$, $\nsf$, and $C_*$, respectively. Column (5)
indicates whether cooling is always on or it is turned off for 40
Myr after SF events in the simulation. The
parameter $\beta$, defined above (see eq. \ref{eq:prob_SF}), 
is shown in column (6). 
We also ran an adiabatic simulation. This 
is denoted in Table~\ref{tab:Models} by the name Ad (see last row).

The SF scheme and the different values of the SF parameters  
$\mlim$, $\nsf$, and $C_*$ 
produce different distributions for the stellar particle masses, 
$m_*$\footnote{
These masses refer to the {\it initial} masses
with which stellar particles were born. At later times, the masses are
always lower than the initial (original) masses because stellar
particles loose mass due to stellar winds and supernova events}.
Figure~\ref{fig:MstarDist} shows this distribution at $z=1.0$ for each
of the runs presented in Table~\ref{tab:Models}. 
Some features can be highlighted from this figure: (a) the 
distributions of the R sequence of runs are wider than the 
corresponding distributions of the D sequence. This just reflects the fact that 
the recipe for SF for the former case is probabilistic; that is,
stellar particles can be produced from a wide range of cold gas
density values. (b) In most runs the distributions have
a spike at their
corresponding $\mlim$ value. (c) From series D, run D4 is the one with 
the largest values of $m_*$. In this run, where
$\mlim$ is set to infinite,  $m_*$ is always equal to 2/3 the gas mass
in the cell. (d) Run R2 has the widest distribution and, along with run R4, 
it shows the largest values of $m_*$. Below, we briefly 
discuss the characteristics of each run. 

Run D1 has $\mlim = 10^4\ \msun$, $\nsf= 50$, and the cooling
is always on (see Table~\ref{tab:Models}). The masses of stellar
particles vary from about $10^4\ \msun$ to $6 \times 10^4 \msun$ and
peak at $10^4\ \msun$. This
run has the highest density threshold and, along with run R4, the
most peaked circular velocity curve (see \S \ref{sec:circ_vel}).  

Run D2 was performed to test the effect of a lower density threshold,
although this necessarily implies that more stellar particles will
form. We thus increased the value of $\mlim$ from $10^4\
\msun$ to $8 \times 10^4\ \msun$ in order to keep the number of stellar
particles computationally tractable. Both of these models have a low $C_*$
value but $\mlim$ is higher in model D2. This means that, 
in practice, in this model, in the first step of the SF
prescription, $m_2 = \mlim$ always (see \S \ref{sec:SF_n_feedback}). Note,
however, that if $m_2 = \mlim$ then the final chosen mass of the stellar
particle cannot be higher than $\mlim$. Thus, in model D2 the masses of
stellar particles are restricted to be lower than $8 \times 10^4\
\msun$. 

Run D3 is similar to D2 but here we turn off the cooling for
40 Myr after SF events. Run D4 is similar to D3 but here
$\mlim = \infty$, which means that 
cells always form stellar particles 
with a 2/3 efficiency factor (see \S \ref{sec:SF_n_feedback}); that is, in any
SF event, 2/3 of the gas mass in the cell is converted into
stars.  The masses of the stellar particles vary from about $10^4\ \msun$ 
to $4 \times 10^5 \msun$.
This model is expected to have a higher impact, as compared with
models D2 and D3, on the  ISM properties of the
simulated galaxy due to the formation of more massive stellar particles,
which are concentrated sources of high energy injection.

Aside from $\mlim$ and $\nsf$, the R sequence of models include the $\beta$
parameter (see eq.\ \ref{eq:prob_SF}). In model R1 this parameter is
100 while in the rest it is 20; model R1 also differs from the rest 
in that it includes ``runaway stars''
\citep{CK2009}, stellar particles able to 
inject energy to the ISM in regions of low-density gas. In this simulation, 
which uses a different version of the code, 
a random velocity, drawn from an exponential distribution
with a characteristic scale of $35 \kms$, is added to the inherited gas
velocity to 30\% of the stellar particles (see Ceverino
\& Klypin 2008 for more details). We used a characteristic velocity twice higher 
than the one used by Ceverino \& Klypin, whose value was motivated by observations, 
to take into account the fact that our simulations are twice less resolved. 
The masses of stellar particles in 
the R runs vary from about $10^3\ \msun$ to $10^6 \msun$. 
Runs R2, R3 and R4 differ only in the
value of $\nsf$. All have $\mlim = 10^4 \msun$, $\beta = 20.0$, and the
cooling switched off for 40 Myr after SF events. They were run to
investigate the effect of varying $\nsf$ on the structural and
evolutionary properties of the simulated galaxy in the stochastic SF
scheme.

\section{Results} \label{sec:results}

As mentioned above, our study focuses on the third most massive halo
found at $z = 1$ because this halo evolves without significant
mergers, at least in the redshfit range from $z\sim 2$ to $\sim 0.4$.  
Most of the analysis of the embedded galaxy is performed at two epochs:
$z = 1$ and 0.43; for the cosmological model used here, the time interval 
between these two redshifts is 3.2 Gyr.

Some global properties of the halo in the adiabatic
simulation, at 
different redshifts, are presented in Table~\ref{tab:AdiaHalo}.
Column (4) shows the concentration of the halo as 
defined by $c_{vir} = \rv/r_s$\footnote{To obtain
$r_s$, the halo DM density profile was fitted 
with the NFW profile \citep{NFW97}.}. The virial mass of the halo 
and the specific angular momentum are given in columns (2)
and (6), respectively. The baryonic fraction within \rv\
is presented in column (3) (see also Table~\ref{tab:BFract}).
Column (4) shows the spin parameter of the halo 
as defined by \citep{Bullock2001}
\begin{equation} 
\lambda = \frac{J}{\sqrt{2} \mv V_{vir} \rv},
\end{equation}
where $J$ is the angular momentum inside the halo virial
radius, and $V_{vir}$ is the circular velocity at \rv. Moreover,
column (2) of Table \ref{tab:BFract} reports the virial masses of the DM halo 
in the other runs. At $z=1$, $M_{\rm vir}$ varies between 4.7 and $5.0 \times 10^{10}\
 \msun$, 
while at $z=0.43$, it varies between 5.4 and $6.6 \times 10^{10}\ \msun$.
In the following, we will discuss the dependence of the 
galaxy morphology, stellar populations, 
dynamics, ISM properties, and SF history of the studied galaxy 
on the relevant parameters of the SF and feedback prescriptions.
\begin{deluxetable}{cccrcc}
\tablecolumns{5}
\tablewidth{0pc}
\tablecaption{Adiabatic Halo Global Parameters}
\tablehead{\colhead{z} & \colhead{$\mv$} & \colhead{$f_{gal}(\rv)$
\tablenotemark{a}} &
\colhead{$c_{vir}$} & \colhead{$\lambda$} & \colhead{J/M \tablenotemark{b}} \\
  & ($10^{10}\ \msun$) &  & &  &  (km s$^{-1}$ kpc)  }
\startdata
  0.0     &   8.6   &  0.14 & 18.4 &  0.044 & 285.1    \\
  0.4     &   5.7   &  0.14 & 13.4 &  0.049 & 212.4    \\
  1.0     &   4.8   &  0.14 &  8.0 &  0.060 & 202.8    \\
\enddata
\tablenotetext{a}{Baryon fraction within the virial radius of the halo.}
\tablenotetext{b}{Specific angular momentum inside \rv.}
\label{tab:AdiaHalo}
\end{deluxetable}

\subsection{Morphology and structure}

The different combinations of the parameters used in the SF and 
feedback prescriptions, described in \S 2.4, produce a large diversity of
galaxy morphologies and stellar populations. As an example,   
in Fig.\ \ref{fig:Diskz=1}
we show the stellar spatial distribution for runs D4 and R2 at $z=1$
and $z=0.43$, with the disk, when present, seen edge-on. 
Stellar particles are color coded according to
their age in a rainbow scale, with the reddest particles being oldest.
Ages range between about 6 Myr to 5.5 Gyr
($z = 1.0$) and 12 Myr to 8.7 Gyr ($z = 0.43)$. 
Unlike run D4, in
which a young stellar disk embedded in an older thick disk can be appreciated, 
the young stellar population of run R2 at $z=1.0$ is distributed in a
``puffy'' spheroidal structure. At $z=0.43$, 3.2 Gyr since $z =1$, 
the once young stellar disk of run D4 has aged, while 
in run R2 a young stellar disk has formed.  This illustrates the large effects  
the SF/feedback prescription has on the structure of the analyzed galaxies
\citep[see also, for example,][]{Scannapieco08}.

 In column (5) of Table \ref{tab:BFract}, we report, for $z=1.0$ and 0.43, the galaxy stellar 
effective radius, \re, defined as the radius where half of the central galaxy 
stellar mass $M_*$ (disk and spheroidal components) is contained. 
We find that the galaxy 
stellar structure is systematically 
larger as $\nsf$ decreases (sequences D1 $\rightarrow$ D2 for the deterministic
model, and R4$\rightarrow$ R3$\rightarrow$ R2 for the random model). This 
radius is particularly small for those runs where feedback is inefficient, 
D1 and D2; when the radiative cooling is artificially switched off allowing 
for thermal pressure--driven feedback, the radius increases by a factor of 
$\sim 2$ (from run D2 to D3). However, \re\ seems to be insensitive to the 
stellar particle mass distribution (compare runs D3 and D4).

\begin{figure}[htb!]
\plotone{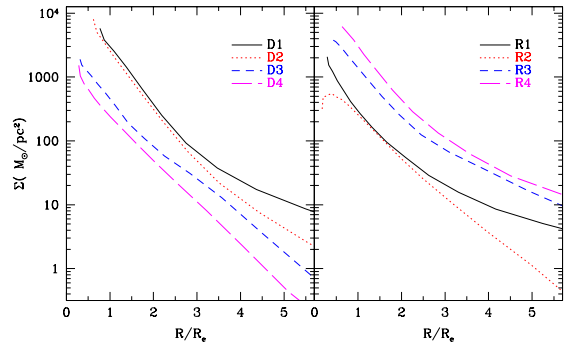}
\caption[fig:SurfDensity]{Stellar surface density profiles for the D 
(left panel) and R sequence of runs at $z=0.43$.
\label{fig:SurfDensity}}
\end{figure}

Figure~\ref{fig:SurfDensity} show the stellar surface density profiles along the disk plane, 
averaged azimutally in concentric cylindrical rings, for all the runs at $z=0.43$;
radii are normalized to their corresponding \re.  In most cases, 
the surface density profiles can be described approximately by two exponential laws: 
an inner nearly exponential profile associated 
to a disk+bulge system and a more extended outer exponential profile associated to 
an older spheroidal structure (stellar halo). It is interesting that the outer
profiles of the R models tend to be higher and more extended than those of the
D models. This is probably because models with a stochastic SF scheme favors 
star formation in low gas density 
environments, helping the formation of an exponential stellar halo.

The measured effective stellar surface densities of the various runs, 
$\Sigma_M = M_*/2\pi R_e^2$, lie in the higher density side but within the range of 
observational determinations at $z\approx 0.4$ and $z \approx 1$; in other words, the 
simulated galaxies have smaller \re\ for their masses than the average radii inferred 
from observations \citep[see e.g.,][]{Barden05}. 

In order to quantify the relative contribution of the spheroidal
and disk components to the total galaxy stellar mass we compute the ratio, 
$\epsilon$, between the
$z$ component of the angular momentum of the stellar particle
and the angular momentum expected 
for a circular orbit, $j_{c} = R V_{c}(R)$. We roughly define the
mass of the spheroid as twice the mass of all stellar particles with
$\epsilon < 0$. The mass of the stellar disk is then estimated as the 
difference between the stellar mass reported in Table \ref{tab:BFract} and the
mass of the spheroid.
In all runs, the spheroidal stellar component is significant (between
$\sim 30$\% and $\sim 60$\% of $M_*$ at $z = 0.43$) and with 
old/intermediate ages. The runs with the best defined stellar disks are R1, D3, 
and R3. The spheroid--to--total 
stellar mass ratio for these runs is $\approx 30-35$\%; the complement is
roughly the disk component. It should be mentioned that run R1 follows a model 
constructed to reproduce Milky--Way--sized galaxies \citep{CK2009}.

\begin{figure}[htb!]
\plotone{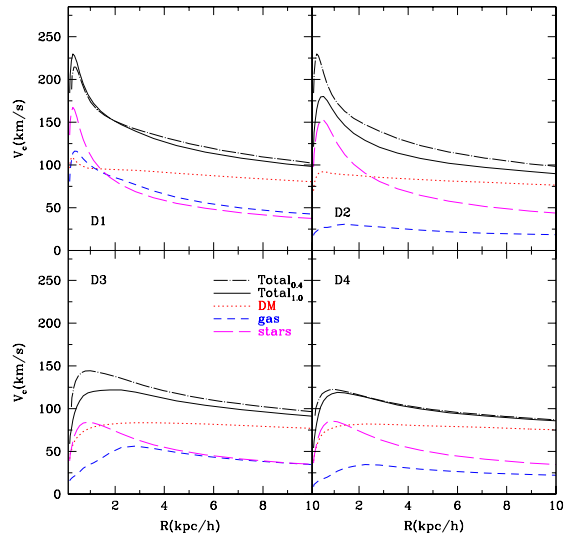}
\caption[fig:VcD]{Circular velocity profile decompostions for the D sequence
of runs at $z=1$: gas (short-dashed line),
dark matter (dotted line), stars (long-dashed line), and total (solid
line). Only the total profiles are shown also for the epoch $z=0.43$ (thick
dot--dashed line).  Runs in which cooling
is always on, D1 and D2, show very peaked profiles, being run D1 the 
most extreme. On the other hand, run D3 and specially run D4, show relatively 
flat profiles.
\label{fig:VcD}}
\end{figure}

\begin{figure}[htb!]
\plotone{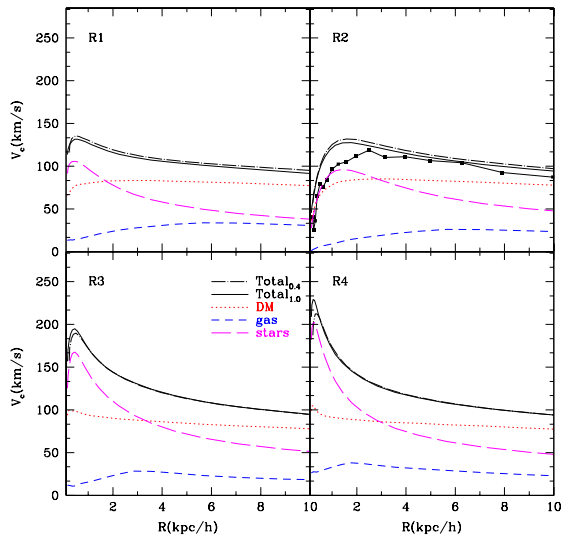}
\caption[fig:VcR]{As in Figure~\ref{fig:VcD} but for the R sequence of runs. Here
we have also added the rotational velocity of the cold gas ($T < 10^4$ K) that is
inside the disk and plotted with a solid line with symbols (squares). 
Runs R2, R3, and R4 differ only in the value of $\nsf$ and it increases 
as we go from R2 to R4. A clear trend of increasing $V_c$ and steepening of
the profile is seen. Runs R2 and R1 have a relatively flat circular 
velocity profile. 
\label{fig:VcR}}
\end{figure}

\subsection{Circular velocity decomposition} \label{sec:circ_vel}

Figures \ref{fig:VcD} and \ref{fig:VcR} show the total circular velocity
profiles for the R and D models described in Table~\ref{tab:Models} at
$z = 1.0$ (solid line) and 0.43 (dot-dashed line). The circular velocity is 
defined as $V_{\rm c} = \sqrt{GM(R)/R}$, where $M(R)$ is the total mass, 
or the mass of a certain galaxy component (dark matter, gas or stars), contained 
within radius $R$. The circular velocities due to the various mass 
components are plotted only for epoch $z=1.0$.

In both, the deterministic (D2 $\rightarrow$ D1) and  
the stochastic (random) runs (R2$\rightarrow$ R3$\rightarrow$ R4), 
we see that an increase in $\nsf$ translates into more centrally concentrated
stellar and gas structures, which produces a more peaked total circular velocity
profile. The halo also becomes more concentrated, but this is a consequence of the
baryonic gravitational drag. As expected, for our spatial resolution,
the contribution of the gas component
to $V_{\rm c}$ increases as $\nsf$ increases. 
This trend might change if tens of parsec resolution is reached.
The lack of efficient feedback produces stellar disks
that are too concentrated (runs D1 and D2); when local cooling is 
switched off temporarily, keeping all other parameters fixed, 
$V_c(R)$ flattens significantly (run D2 to D3).

It is interesting to note that run R2 has a mild
declining circular velocity curve. However, this run has no disk at $z = 1$, as 
seen in Figure~\ref{fig:Diskz=1}.  Nevertheless, a disk forms later on and 
its $V_{\rm c}$ becomes sliglthly more peaked.
Runs D3 and D4 also have mild circular velocity curves at 
$z=1$, but, unlike run R2, these runs do have well defined young stellar 
disks at this redshift.
As time goes on, the circular velocities of these models become more peaked 
and the galaxies more concentrated (see Figure \ref{fig:VcD}).
Interestingly enough, such an evolution is much less pronounced for runs 
with the random SF prescription (see Figure \ref{fig:VcR}).

Lastly, it is important to say  that the average rotational velocity in the mid-plane of
the disk does not trace the galaxy circular velocity
in our simulated galaxies (see, 
for example, Figure \ref{fig:VcR}). 
In some of our models, the tangential velocity profile 
slowly increases with radius as it is observed in low-mass galaxies  
\citep[\eg][ but see Swaters et al. 2009]{Persic96}. Notice, however,
that at the current resolution our galaxies do not present a central bar. 
Similar situations in which rotation curves do not trace the circular velocity 
have been discussed by \citet{Valenzuela07} and \citet{Rhee04}. In these cases 
the difference is attributed to pressure gradients and non-circular motions.

\subsection{Star formation history}

\begin{figure}[htb!]
\plotone{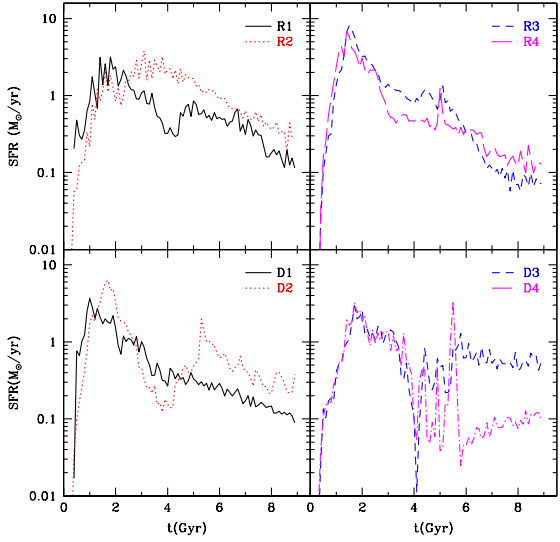}
\caption[fig:SFH]{SFR histories for all runs presented in this paper. 
Each panel shows two runs labeled inside panels. In the X axis
runs the cosmic time. The last time shown corresponds to $z=0.43$. For
given run, the SFR as a
function of time is computed using the snapshot at $z = 0.43$ by tracing back 
the recorded formation times and 
initial masses of the stellar 
particles that are inside the virial radius. 
\label{fig:SFH}}
\end{figure}

Figure~\ref{fig:SFH} shows the SF
history (SFH) for the various runs, defined as the instantaneous 
SFR as a function of time, and computed for
each run using the last snapshot recorded. Specifically, in each data
dump, in addition to the positions and velocities for all stellar
particles, we also save the time at which they formed, their masses (initial
and present), and metallicities due to supernovae of types Ia and II.  
We add up the (initial) masses of the stellar particles
formed during a certain time interval, which we take as 0.1 Gyr, and
divide it by this time, to obtain the ``instantaneous'' SFR. 
Since the amount of stellar mass outside the
galaxy is small during the epochs analyzed ($<$ 3\%), the SFH computed
inside \rv\ is basically that of the galaxy. The simulations were run
until $z=0.43$, whichs corresponds for the cosmology used here to a cosmic age of
8.9 Gyr.

In all runs, the SFHs are characterized by a fast increase lasting 
in most cases 1.5--2.0 Gyr followed by a nearly exponential decreasing phase.  
Runs with smaller values of $\nsf$ show a more gradual decrease in general and the
peak of the SFH is shifted to later epochs (sequences R4$\rightarrow$ R3$\rightarrow$ R2, 
and D1$\rightarrow$ D2). In the case of the R sequence, the increasing phase is also 
more gradual and the amplitude of the maximum decreases with decreasing $\nsf$. 
Run R2, which has $\nsf = 0.1\ \pcc$, exhibits the most gradual SFH 
initial increase and the 
latest SFH maximum ($t \sim 3.5$--4 Gyr). Runs R3 and R4
are those with the fastest SFR decline with time.

The SFH is also stronlgy affected by the stellar 
ability to heat the surrounding gas. The amplitude of the initial 
maximum in run D3, in which the cooling is temporarily
switched off, is lower, and its subsequent 
decrease is more gradual, as compared with run D2. 
The SFH of the former run is much burstier 
than the one of the latter run, with periods of very low SF activity followed 
by intense bursts. This bursty behaviour is even more pronounced in run D4, 
a run similar to D3 but with more massive stellar particles
(compare also the SFHs of runs D1 and D2). 
Notice also that in general runs with the random SF prescription tend to have a 
less bursty SFH than those with the deterministic one.

\subsubsection{Relationship between star formation rate and gas density}

\begin{figure}[htb!]
\plotone{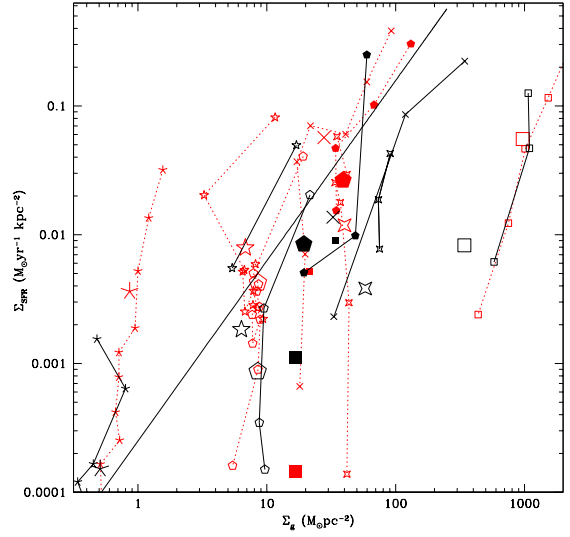}
\caption[fig:SigmaSFR]{Radial SFR surface densities, \sigSFR, versus radial gas surface 
densities, \siggas, for each one of the runs at two epochs. The points of a given run
are plotted with the same symbol and connected with (red) dotted lines for the epoch
$z=1.0$ and with solid (black) lines for $z=0.43$. The large symbols show the corresponding
average value of \sigSFR\ and \siggas\ of each run at the two epochs (see the text for more 
details on how the radial and average \sigSFR\ and \siggas\ were calculated ). The 
symbol code for each run is as follows: D1 (open square), D2 (skeletal, four-end),
D3 (open star, four-end), D4 (solid square), R1 (open pentagon), R2 (skeletal,
five-end), R3 (open star, five-end), and R4 (solid pentagon). The solid line
is the fit to observations by \citet{Kennicutt98}.
\label{fig:SigmaSFR}}
\end{figure}

Figure \ref{fig:SigmaSFR} shows the 
average SF rate (SFR) surface density, \sigSFR, versus 
the gas surface density, \siggas, for all the runs at $z = 1$ and $z=0.43$.
The surface SFRs are computed using the surface densities of stellar particles with ages
smaller than $ 4.0 \times 10^7 \yr$.  For the gas surface density,
only gas with $T_g< 10^4$ K is considered. Cylindrically-averaged values 
of both quantitites with a constant radial bin of 0.5\re\ wide 
and 0.5 \kpch\ proper height are obtained for each model and each epoch. 
For each run, a number of points are plotted with this number 
depending on the extent 
of the simulated galaxy. The symbols corresponding to each run are given in the
figure caption. The symbols connected by the dotted and solid lines correspond to
$z=1.0$ and $z=0.43$, respectively. The large symbols correspond to averages
of \sigSFR\ and \siggas\ inside 3\re.  Notice that surface gas densities move to 
the right (left) if we take a longer (shorter) cylinder.

As expected, as we go to smaller $\nsf$ values (sequences R4$\rightarrow$ R3$\rightarrow$ 
R2, and D1$\rightarrow$ D2), the points 
move down and to the left side of  the \sigSFR--\siggas\ 
diagram. Stellar feedback (see for example, from run D2 to run D3), does not impact 
significantly the \sigSFR--\siggas\ diagram; at $z=0.43$
these runs occupy a similar locus in this diagram.
As time goes on, while \siggas\ decreases
slightly, \sigSFR\ decreases strongly for most of the cases. In general, 
for a given simulation, the correlation of \sigSFR\ with \siggas, both along 
the radius and at two epochs, is steep and becomes steeper as \siggas\ decreases.

\subsection{Properties of the interstellar medium} \label{sec:Props_ISM}

\subsubsection{Probability density functions (PDFs)} \label{sec:PDFs}

\begin{figure}[htb!]
\plotone{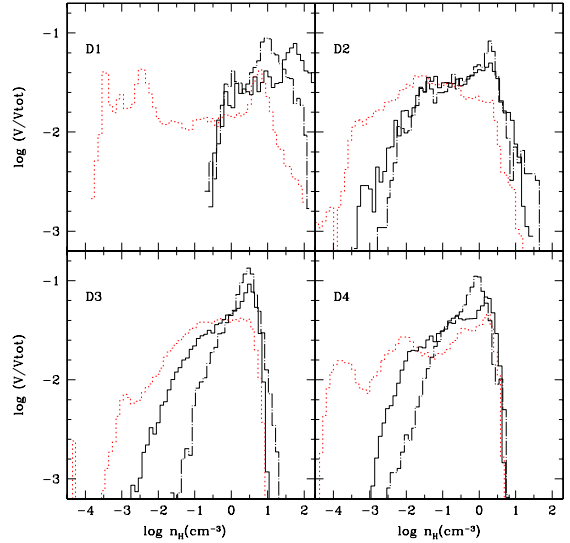}
\caption[fig:PDFdenD]{Probability distribution functions of gas density, $\nH$,
for the runs from series D. The $\nH$ PDF's, the fraction of
total volumen occupied by gas in a given density logarithmic bin, were
computed at $z = 1$ or 0.43 using all cells in the maximum level of refinement (dotted
lines) or using the cells contained in a cylinder of radius 3\re\ and
height 0.6 \kpch\ proper (solid and thick dot-dashed lines) centered on 
the disk. At $z = 0.43$ only the PDFs within the cylinder 
(thick dot-dashed lines) are shown
in the figure to avoid line crowding.
The PDF's computed using the maximum level of refinement 
cover a bigger volume including a greater normalized
fraction of low-density gas.
\label{fig:PDFdenD}}
\end{figure}

\begin{figure}[htb!]
\plotone{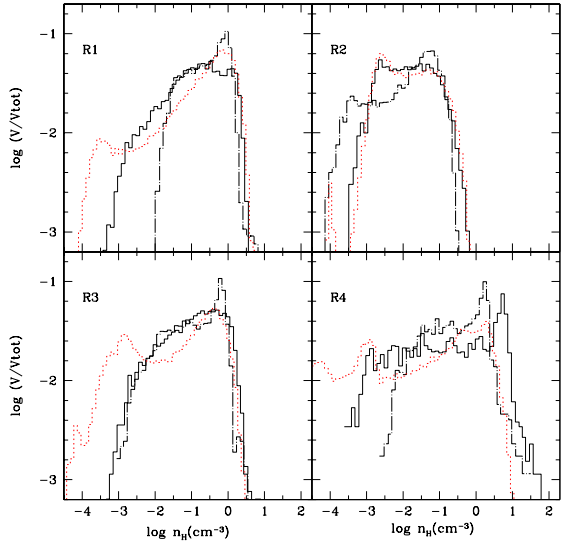}
\caption[fig:PDFdenR]{As in Figure~\ref{fig:PDFdenD} but for the models of the
series R. 
\label{fig:PDFdenR}}
\end{figure}

Figures \ref{fig:PDFdenD} and \ref{fig:PDFdenR} show the
normalized probability density function, PDF, of gas number density for
the D and R series, respectively. Thin solid and dotted lines refer to 
measurements at $z=1$. The latter show the PDFs computed using all cells 
in the maximum level of refinement, $l = \lmax$, while the former  
represent the PDFs computed using the cells contained within a cylinder 
with a radius of 3\re\ and height of 0.6 \kpch\ proper,
centered on and oriented with the disk. The normalization volume, $\Vtot$,
differs in both cases, being smaller for the cylinder. The thick dot-dashed
line refers to the PDF measured for the gas within the cylinder at $z=0.43$. 
The plane of the disk is defined as the plane normal to the angular momentum 
vector of the cold gas located in an inner cylindrical shell of radii of 0.5 
and 3 \kpch\ comoving.  We have varied the inner and outer radii of the cylinder 
and found negligible variations in the measured quantities.
Notice that PDFs computed using the cells at $l = \lmax$ 
include a significant fraction of cells that contain low-density warm and hot gas. 
This is due to the fact that these low-density gas cells are kept refined 
because the dark matter density cell refinement criterium is satisfied.

\begin{figure}[htb!]
\plotone{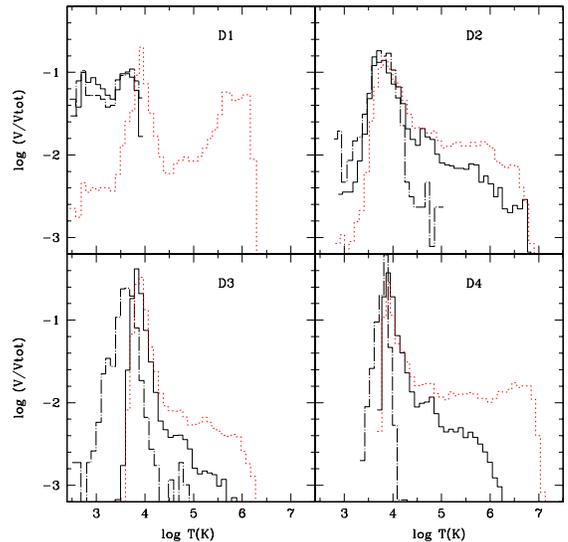}
\caption[fig:PDFtempD]{Probability distribution function of gas temperature, $T$, 
for the runs from series D. The PDF's of temperature, the fraction of
total volume occupied by gas in a given temperature logaritmic bin, were
computed as explained in the text and in Figure \ref{fig:PDFdenD}. The same line
code as in Figure \ref{fig:PDFdenD} is used here. 
\label{fig:PDFtempD}}
\end{figure}

\begin{figure}[htb!]
\plotone{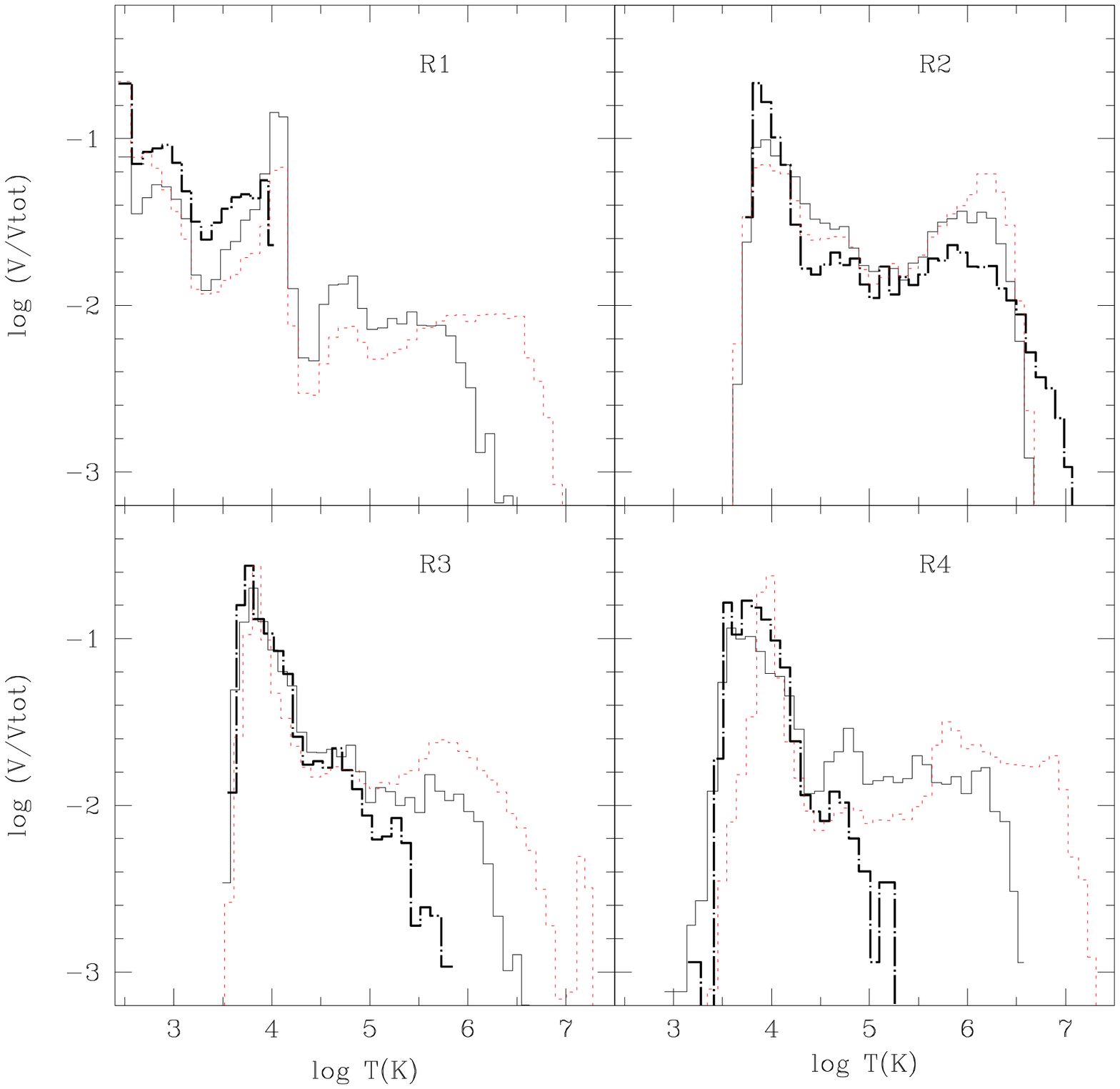}
\caption[fig:PDFtempR]{As in Figure~\ref{fig:PDFtempD} but for runs of
series R. 
\label{fig:PDFtempR}}
\end{figure}

Figures ~\ref{fig:PDFtempD} and \ref{fig:PDFtempR} show the
corresponding PDFs of the gas temperature; the line code is as in
Figures \ref{fig:PDFdenD} and \ref{fig:PDFdenR}. As was already noticed, the
volume occupied by the cylinder is smaller than that of the cells at the
maximum level of refinement. In particular, the temperature PDF
of run D1 computed over the cylinder at $z=1$ contains only cells with 
$T < 10^4$ K (see upper-left panel of Figure~\ref{fig:PDFtempD}); 
that is, the PDF of this model is only sampling the well-defined cold 
gaseous disk.  Instead, the PDFs computed using the cells at $l = \lmax$ contain 
a significant fraction of warm-hot/hot gas.

Figures \ref{fig:PDFdenD}--\ref{fig:PDFtempR} show that at $z=1.0$ 
(the situation is similar at $z=0.43$, albeit more extreme) all models
except R2 have very different PDF's when calculated over the cylinder
around the mid-plane than when computed using the
grid cells at $l = l_{max}$. The main difference is the relative
absence of hot, low-density gas in the mid-plane: this gas occupies
a very small fraction of the disk volume. In particular, we see 
that in runs D3 and D4, the maximum temperatures in the disk are 
$T \sim 10^{5.5-6}$ K. On the other hand, we see that the 
temperature PDF of the finest grid of run D4, in particular, contains a 
large amount of hot gas, and that its distribution reaches 
temperatures of up to $10^7$ K. This suggests that the hot gas produced by stellar
feedback in the disk of run D4 becomes buoyant and leaves the
disk. Indeed, run D4 even exhibits a wind (cf.\ \S \ref{sec:ff_prof}). 

Run R2 is the only one at $z=1$ for which the density and temperature
PDF's are very similar when measured for the mid-plane cylinder and for
the finest grid. This can be understood because the galaxy in this run 
has a more spheroidal shape at this redshift. 
However, at $z=0.43$ (thick dot--dashed line) a thick disk
is already in place (see Figure~\ref{fig:Diskz=1}) and the PDF's measured 
for the mid-plane cylinder and for
the finest grid start to deviate from each other, with the low--density hot disk 
gas buoyantly escaping to the halo. 

Let us analize the PDFs calculated over the cylinder around the rotating
mid-plane. We see that as $\nsf$ is made smaller
(sequences R4$\rightarrow$ R3$\rightarrow$ R2, and
D1$\rightarrow$ D2), the density and temperature PDFs
become broader. 
The main effect of the broadening is in the direction of increasing the 
amount of low--density and warm--hot/hot gas. In the extreme 
case of model R2, at $z = 1$ a second peak appears in the low-density 
and high-temperature sides of the corresponding PDFs (see Figures \ref{fig:PDFdenR} 
and \ref{fig:PDFtempR}).
The maximum gas density reached by a run also depends on the value 
of $\nsf$: the lower $\nsf$, the lower the maximum $\nH$ value. 

Run D2, in comparison to D3 (Figures \ref{fig:PDFdenD} and \ref{fig:PDFtempD}),
shows a higher fraction of low-density, hot gas in both the cylinder
or in the highest-refined grid cells
(dotted lines). We see here
that a stronger feedback (run D3) produces a less dominant hot gas phase, very 
likely because part of this hot gas is expelled from the mid-plane of the disk.
In any case, the galaxy in run D2 is too compact (see Table \ref{tab:BFract}) and shows 
a rather thick young stellar disk, with its gaseous structure 
resembling that of run R2. The region in which we compute
the PDFs contains a mixture of cold and warm-hot/hot gas, heated very likely by 
shocks.

From run D4 to run D3 (Figures \ref{fig:PDFdenD} and \ref{fig:PDFtempD}), 
the PDFs of density and temperature both within the disk at $z = 1$
(thin solid lines) and in the cells at the maximum refinement level  
(dotted lines) show a decrease in the lower density and higher 
temperature regimes.
Therefore, by allowing for more massive stellar particles
(a more efficient conversion of gas to stars), the galaxy produces
a higher fraction of low-density and hot gas.

\subsubsection{Vertical filling--factor profiles} \label{sec:ff_prof}

\begin{figure}[htb!]
\plotone{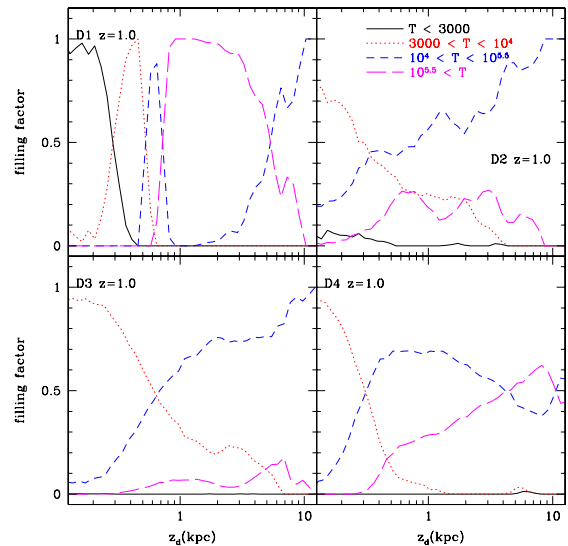}
\caption[fig:FFD]{Fraction of volume filled with gas at different temperatures
as a function of height (below and above the disk plane) for runs of series 
D at $z= 1$. All runs of the series D present no
hot phase in the mid-plane of the disk ($|z| < 300$ pc/h) while runs from 
D2 to D4 practically have no cold gas ($T < 3000$ K). 
On the other hand, run D4 is the only D model that has a fraction
of hot gas that increases as we go away from the mid-plane of the disk.
\label{fig:FFD}}
\end{figure}

\begin{figure}[htb!]
\plotone{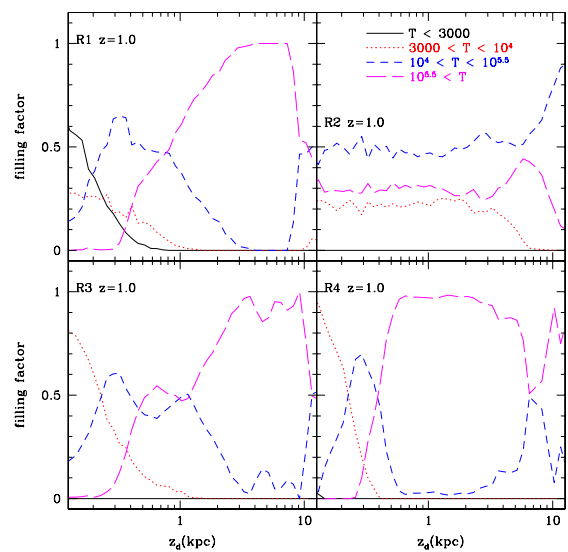}
\caption[fig:FFR]{As in Figure~\ref{fig:FFD} but for runs of series R. 
As it was the case for runs D2 to D4, runs R2 to R4 have practically
no cold gas whereas, as it happens with run D4, here runs R1 and R3
have a fraction of hot gas that increases as we go away from the
mid-plane of the disk. On the other hand, run R2 is the only one in
which we found a non-negligible fraction of hot gas in the mid-plane of
the disk.
\label{fig:FFR}}
\end{figure}

The spatial distribution of the gas in the runs can also be analysed
through the filling factors of the gas at various temperature
ranges. The vertical profiles at $z = 1$ are shown in Figures \ref{fig:FFD} 
and \ref{fig:FFR}. 
We compute the filling factor over cylinders that have the
same radius as the one we use to compute the PDFs; i.e.,
3\re\ of the corresponding simulated galaxy.
 We define four regimes according to the (azimuthally averaged)
gas temperature: cold gas ($T < 3000$ K, {\it solid lines}), warm 
gas ($ 3000 \leq T(\Ke) < 10^4$, {\it dotted lines}), warm-hot gas ($
10^4 \leq T(\Ke) < 10^{5.5}$, {\it short-dashed lines} ), and hot gas
($T \ge 10^{5.5}$ K, {\it long-dashed lines}).

Figures \ref{fig:FFD} and \ref{fig:FFR} show that as $\nsf$ decreases 
(sequences R4$\rightarrow$ R3$\rightarrow$ R2, and D1$\rightarrow$ D2)
and/or $\mlim$\ increases (sequence D1$\rightarrow$ D2)
the fraction of colder gas in the mid-plane decreases and
the fraction of hotter gas increases. From these two figures the 
following other features can also be highlighted: (a) the filling factors 
of the different gas components change strongly with the height above 
and below the mid-plane, $\vert z_d \vert$. Typically, the fraction of 
colder gas decreases and that of the hotter gas increases with $\vert z_d \vert$; 
(b) run R2 is the only one where these filling factors remain nearly constant
with $\vert z_d \vert$, and it is also the only case that presents a non-negligible fraction 
of hot gas at low disk heights; (c) the ISM in the 
disk mid-plane of run D1 is composed only of the cold phase
; (d) a multiphase ISM with a non-negligible fraction of cold gas at low heights 
exists only for run R1.

As can be seen in Figs. \ref{fig:FFD} and \ref{fig:FFR},
the only two runs that presents cold gas 
in the mid-plane are run D1 and R1. The mid-plane of run D1 is 
dominated by cold gas because it has a high $\nsf$ value,
a low stellar mass limit $\mlim $, and an inefficient stellar 
feedback. Run R1, on the other hand, has a
significant amount of cold gas in the mid-plane
because in this run the UV background ionizing
radiation was artificially reduced\footnote{The formula 
for the ionizing flux, which is based
on \citet{HM96}, is taken from \citet{KravTh}. For reference, we note
that the fluxes at redshifts $z=8,\ z_{\rm max}$, and 1.0 are $5.9 \times 10^{-26}$,  
$7.2 \times 10^{-22}$, and  $8.5 \times 10^{-23}$ erg s$^{-1}$ cm$^{-2}$ Hz$^{-1}$,
respectively, where $z_{\rm max}$ denotes the redshift at which the
ionizing flux is maximum.} from $z = 8$ to present-day to its
value at $z = 8$ and also because it has the largest
$\beta$, implying that some of the cold dense gas does not form stars,
and therefore can remain as such.

\begin{figure}[htb!]
\plotone{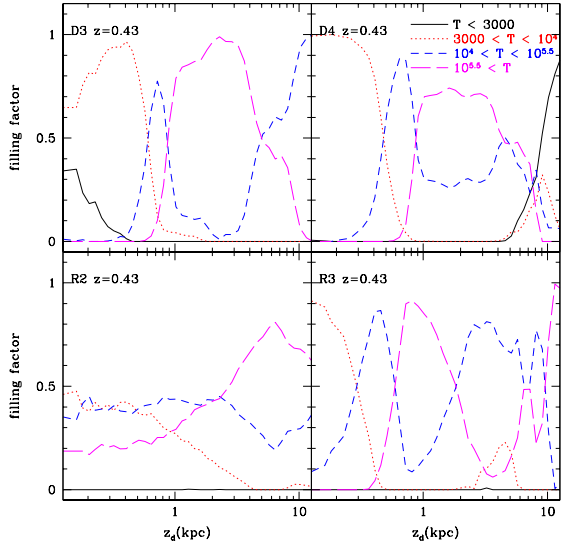}
\caption[fig:FFlater]{Fraction of volume filled with gas at different temperatures
as a function of height (below and above the disk plane) for runs D3, D4, R2, and R3
at $z= 0.43$. These are the runs that most change their vertical filling factors since
$z=1$.
\label{fig:FFlater}}
\end{figure}

In Figure \ref{fig:FFlater} we plot the vertical filling factor profiles of runs 
D3, D4, R2, and R3 at $z=0.43$ in order to see how much these profiles could
change with time. These are the cases that change most from
$z=1$ to $z=0.43$. The general trend is that the fraction of colder gas 
increases in the mid-plane with time (specially in run R3), while the hotter gas 
`moves' to higher vertical distances. In most of the runs, one observes that with 
time the warm gas fraction accumulates in an narrow extraplanar layer at 
$\sim 0.4-0.8$ kpc that coexists with the hot gas. At larger
heights, the fraction of hot gas already dominates. 
Run D3 at $z=0.43$
has at the mid-plane $\sim 20\%$ and $\sim 80\%$ of the volume occupied by cold 
and warm gas, respectively, while in run D4, $\sim 100\%$ is occupied by 
warm gas. The filling factor of the warm-hot gas in both runs peaks at a height 
$\sim 700$ pc, at $\sim 75$\% and $\sim 90$\%, respectively. The hot gas becomes 
the dominant component at a height $\sim 1$ kpc. For the runs R3 and R2, 
the warm-hot gas is present in the mid-plane, with volume fractions of $\sim 20\%$
and $\sim 40\%$, respectively. 
 
\begin{deluxetable*}{ccccccccc}
\tablecolumns{9}
\tablecaption{Global properties of models}
\tablehead{\colhead{Model} & \colhead{$\mv$} &  \colhead{$M_*$} &  \colhead{$f_g$
\tablenotemark{a}} & \colhead{$\re$\tablenotemark{b}} & 
\colhead{$f_{gal}(\rv)$\tablenotemark{c}} & \colhead{$f_{gal}(5\re)$\tablenotemark{d}} & 
\colhead{J/M(halo)} & \colhead{J/M(baryons)} \\
   & ($10^{10}\ \msun$) & ($10^{10}\ \msun$) &  & (kpc) &  &  & (km s$^{-1}$ kpc) & (km s$^{-1}$ kpc) }
\startdata
 \multicolumn{9}{c}{$z = 1$} \\
  D1     &  4.69    &   0.312  &  0.412  &  0.24 & 0.16  & 0.10 &  96.0  &  88.3 \\
  D2     &  4.69    &   0.466  &  0.070  &  0.49 & 0.12  & 0.10 & 203.9  &  59.9 \\
  D3     &  4.87    &   0.308  &  0.446  &  1.16 & 0.13  & 0.11 & 208.8  & 287.1 \\
  D4     &  4.87    &   0.304  &  0.210  &  1.17 & 0.10  & 0.08 & 195.9  & 143.3 \\
  R1     &  4.79    &   0.332  &  0.133  &  0.90 & 0.13  & 0.07 & 188.1  & 118.6 \\
  R2     &  4.99    &   0.579  &  0.076  &  1.46 & 0.14  & 0.11 & 205.2  & 107.6 \\
  R3     &  4.81    &   0.625  &  0.043  &  0.64 & 0.15  & 0.12 & 218.9  &  90.5 \\
  R4     &  4.80    &   0.524  &  0.065  &  0.37 & 0.15  & 0.10 & 232.0  &  51.3 \\
 \multicolumn{9}{c}{$z = 0.43$} \\
  D1     &  5.40    &   0.358  &  0.396  &  0.41 & 0.17  & 0.10 & 143.9  & 119.5 \\
  D2     &  5.54    &   0.546  &  0.060  &  0.51 & 0.13  & 0.10 & 218.1  &  57.9 \\
  D3     &  5.84    &   0.429  &  0.369  &  1.00 & 0.13  & 0.11 & 199.6  & 261.3 \\
  D4     &  6.60    &   0.309  &  0.229  &  1.11 & 0.09  & 0.06 &  49.4  & 149.7 \\
  R1     &  5.64    &   0.379  &  0.161  &  1.08 & 0.13  & 0.07 & 240.8  & 159.1 \\
  R2     &  6.21    &   0.676  &  0.075  &  1.80 & 0.13  & 0.11 & 141.2  & 167.8 \\
  R3     &  5.49    &   0.640  &  0.043  &  0.73 & 0.13  & 0.11 & 203.7  &  93.8 \\
  R4     &  5.63    &   0.571  &  0.051  &  0.50 & 0.13  & 0.10 & 229.4  &  49.9 \\
\enddata
\tablenotetext{a}{Gas fraction as defined by $f_g \equiv M_{\rm gas}/(M_{\rm gas} + M_*)$, where
only gas cells with $T_g < 10^4$ K are used.}
\tablenotetext{b}{Effective radius defined as the radius where half of the central galaxy stellar 
mass is contained.}
\tablenotetext{c}{Baryon fraction within the virial radius of the halo using all baryons, gas
at all temperatures plus stars.}
\tablenotetext{d}{Baryon fraction within five effective radius using only 
stars plus gas with $T_g< 10^4$ K.}
\label{tab:BFract}
\end{deluxetable*}

\subsection{The mass and angular momentum fractions} 

In Table \ref{tab:BFract}, $M_*$ and the gas fraction, $f_g$,
of the galaxies in the different runs are reported for $z=1$ and 0.43. 
The gas fraction is defined as $f_g \equiv M_{\rm gas}/(M_{\rm gas} + M_*)$, where
only gas cells with $T < 10^4$ K are used for the gas.  The stellar and gas masses 
were measured within a sphere of radius 5\re. We see that for runs without efficient 
feedback, runs D1 and D2, $f_g$ strongly decreases (gas is consumed efficiently) 
as $\nsf$ diminishes and, to a lesser degree, as $\mlim$ increases.
Runs with a stochastic SF scheme do not show a clear trend of
$f_g$ with $\nsf$. From run R4 to R2, $f_g$ first decreases (runs R4$\rightarrow$ 
R3) and then it increases (runs R3$\rightarrow$ R2). On the other hand,
run R1 is the run, within the R series of runs, with highest $f_g$ fraction
at both $z = 1$ and 0.43.

The shape of stellar particle mass distribution seems to influence 
the value of $f_g$: 
for cases where this distribution extends to high-mass stellar particles, $f_g$ is 
smaller (from run D3 to D4 and partially from run D1 to D2). Note that in run D4 
the feedback is so strong that a galactic wind is produced and as
a result a signicant fraction of gas is ejected from the disk. 
Finally, from Table \ref{tab:BFract}
it is inferred that runs with the random SF scheme are more efficient in transforming gas 
into stars than those with the deterministic scheme (the $f_g$ values of 
R runs are 
systematically smaller than those of runs D). 
However, it should be noted that as a consequence of the low value of $C_*$ 
and the SF scheme, gas is transformed locally into stars less efficiently in runs
D1 to D3 than in the R runs.

Columns (6) and (7) of Table \ref{tab:BFract} show the mass baryon fraction
using all baryons (gas at all temperatures + stars) within \rv, 
and using only those baryons (gas with $T_g< 10^4$ K + stars) within a sphere of 5\re\ around
the galaxy center, respectively. The latter case is related 
to what is called usually the galaxy baryon fraction. Several kind of estimates 
and observational inferences show that this fraction in present--day galaxies is 
much smaller than the universal value, equal to 0.15 
for the cosmology used here. A radius of up to 5\re\ was used to compute $f_{gal}$
to encompass the gaseous disk, which is typically larger than the stellar
disk.

The galaxy mass baryon fraction, $f_{gal}(5\re)$, does not depend on the value of $\nsf$,
as is seen from the runs where this parameter is changed systematically,
(sequences R4$\rightarrow$ R3$\rightarrow$ R2 and D1$\rightarrow$ D2).  
From run D2 to D3, $f_{gal}(5\re)$ is almost the same, showing that switching off
local cooling by 40 Myr, at least in this simulation, does not produce 
a net loss of baryons from the galaxy. However, from run D3 to D4, 
both the halo and galaxy baryon fractions decrease significantly. At $z=0.43$ run
D4 has the lowest values of these fractions among all the runs. 
Therefore, a bias toward massive particles $m_*$ distribution
along with efficient feedback induces strong gas ejection
from the galaxy and halo.

According to Table \ref{tab:BFract}, the fraction of baryons within the halo
but outside the galaxy is non-negligible for all the runs. The extreme cases
are runs D1, R1 and R4 where 50\% or more of the baryonic content of the 
corresponding halo is outside the central galaxy. These baryons are in the form
of shock--heated gas or gas heated and expelled from the central galaxy; a small
fraction is in stars in satellite galaxies. 
The halo with the smallest baryon fraction corresponds to run D4 at $z= 0.43$,
with $\sim 60\%$ of the universal fraction. This run is the one with the most
efficient galaxy outflows.

Columns (8) and (9) of Table \ref{tab:BFract} present the total
specific angular momentum (AM) for the dark matter halo and for the sum
of the stellar and cold gas components of the central galaxy 
(within a sphere of 5\re\ radius), respectively. The specific
AM is given in units of km s$^{-1}$ kpc.
The total specific AM measured at a given time is sensitive to 
the current halo and galaxy assembly events, nevertheless, with a couple
of exceptions, values are around 200 km s$^{-1}$ kpc, a value similar to 
those found in the adiabatic run (see Table~\ref{tab:AdiaHalo}). In this
latter case the halo is almost not `perturbed' by the baryonic processes. 
We found that the specific AM of the DM halo is typcally, as
expected, greater than
the corresponding specific AM of the baryonic component.
The effects of complex baryonic processes in the different runs 
produce variations in the halo mass assembly 
history as well as in the properties of the surviving subhalos. 
Such variations are on the
basis of the different values of the halo's specific AM measured 
in our different runs, at epochs where the halo is 
yet in a relatively active phase of mass aggregation.

In spite of the transient nature of the specific AM measures 
mentioned above, some 
general results can be highlighted. We see, as expected, 
that in most of the runs 
the halo's specific AM is larger than the galaxy's specific AM.  
Nevertheless, the galaxy's specific AM for those runs with
strong feedback is relatively large and in some cases 
it is even larger than the corresponding halo's specific 
AM (see Table~\ref{tab:BFract} run D3). The 
specific AM in most of the runs remains roughly constant
or slightly increases from $z=1$ 
to $z=0.43$. The effect of feedback is clearly in the direction of 
producing galaxies with larger specific AM (see the change from run D2 to D3).
We also see that the galaxy's specific AM increases as $\nsf$ decreases 
(sequence 
R4$\rightarrow$ R3$\rightarrow$ R2), but this seems to be related to 
strength of the feedback, it is stronger in low gas density.

\begin{figure*}
\includegraphics[width=\textwidth]{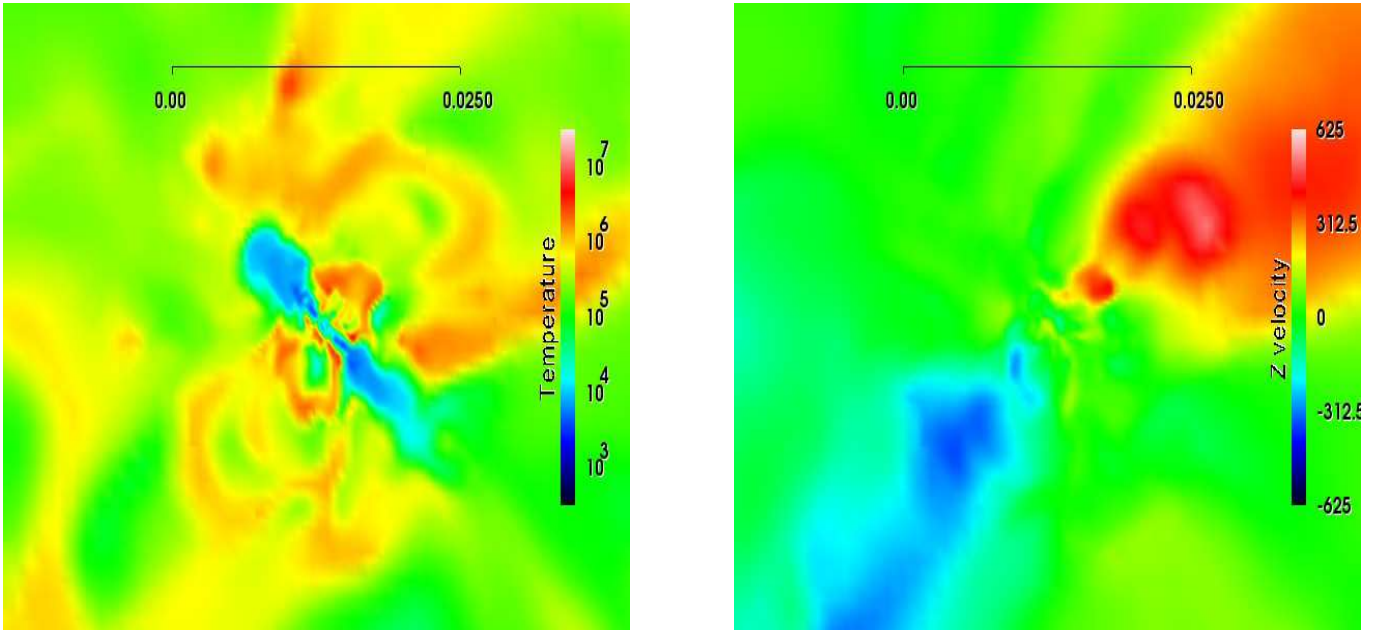}
\caption[fig:VzImage]{Edge-on view of simulation
R3 at $z = 1$. {\it Left panel:} Distribution 
of the gas temperature. {\it Right panel:} Distribution of the $z$
velocity component. The slices are 
36 comoving kpc on a side 
(length of solid line) and 219 comoving pc thick. It can be
seen that gas velocities can clearly exceed 200 $\kms$.
\label{fig:VzImage}}
\end{figure*}

\section{Discussion}

\subsection{General trends and SF/ISM properties}

In the sequences of runs D2 $\rightarrow$ D1 
(deterministic SF model with weak feedback) and R2$\rightarrow$ R3$\rightarrow$ R4, 
(random SF model with efficient feedback) {\it the SF density threshold $\nsf$} is increased, 
keeping the other parameters 
similar. Our simulations show that as $\nsf$ is increased, the overall stellar galaxy 
has a smaller effective radius \re\ and the circular
velocity profile is more peaked. For lower values
of $\nsf$, the peak in the SFH tends to be slightly shifted to later epochs and the 
later (nearly exponential) decline in SFR is more gradual. Related to this trend, 
at smaller $\nsf$, the stellar populations become younger
on average, the stellar structure becomes thicker
\citep[see also][]{Saitoh09}, and a larger fraction of stars exists in a ``puffier'',
more spheroidal component.

Regarding the ISM properties of the simulated galaxies, their PDFs
of the density and temperature 
evolve to broader distributions, with more gas in the low-density,
and high-temperature regime in the disk, 
and a smaller maximum value for the gas density, 
as $\nsf$ is smaller. The gas filling factors 
above and below the mid-plane also show that for 
smaller $\nsf$, the cold gas close to the mid-plane
become less abundant. We note that in general is 
not easy to produce cold, high--density gas in the simulations. At $z=1.0$, only 
the run with the highest $\nsf$ ($=50\ \pcc$) and weak feedback (run D1) 
shows a significant amount of this gas in the mid-plane. At $z=0.43$, cold and
high--density gas is also present in the mid-plane of run D3. 
On the other hand, a hot gas phase in the disk
mid-plane can only apparently be produced by lowering $\nsf$.

It is not clear what a realistic value for $\nsf$ should be,
but we argued in \S 2.2
that for a cell of 150 pc, close to the 218 pc of our nominal resolution
at $z = 0$, this value should be around 5 $\pcc$. Moreover, the results of
our experiments seem to support this number: a value much higher than this
produces a galaxy that is too compact (see, 
for example, R2 and R3 versus R4
or D2 versus D1, albeit with weak feedbak), while a much lower value produces
a system that is closer to being a spheroidal galaxy, with a very small 
fraction of gas, rather than a disk (see Table \ref{tab:BFract}). However, a small value of 
$\nsf$ (0.1 $\pcc$) may produce
reasonable results as shown, for example, by \citet{Governato07}. We speculate 
that this may be an effect of the different technique (AMR versus SPH) and 
subgrid physics (feedback) used to simulate galaxy formation.

\citet{Saitoh09} concluded that values of 
$\nsf$ as high as $\sim 100\ \pcc$ should be used in high--resolution 
simulations ($\sim 20$ pc) in order
to reproduce the complex structure of the gas disk. They also concluded
that using high values for $\nsf$ makes the model results fairly insensitive to the 
SF prescription. In our experiments, high values of $\nsf$ produce systematically too 
concentrated galaxies with the hot gas phase being absent from the mid-plane. Yet,
results in \citet{Saitoh09} are actually in agreement with our suggested strategy for
choosing $\nsf$ on the basis of the gas column density. For a 20-10 pc 
resolution, our simple argument would indeed suggest values of $\nsf$ of 
50-100 $\pcc$

A second parameter that enters in the SF prescription is the 
{\it stellar particle mass limit $\mlim$}, defined in \S 2.2. As was mentioned in
that section this is not an absolute lower limit. 
High values of $\mst$, for a given cell size, mean that
more stellar and SN energy is dumped into the cell. As a result, in models
with cooling turned off in the SF regions, the feedback--driven ejection of 
gas from the disk increases and the SFH of the 
modeled galaxy becomes more bursty (compare runs D3 and D4). On the other hand, the 
structural/dynamical properties of the modeled galaxies does not seem to depend strongly 
on the stellar particle mass distribution.

Regarding the {\it efficiency of stellar feedback}, we find that it has a strong effect on 
the structural/dynamical properties of the galaxy as well as on their 
SFH and ISM properties, in agreement with recent numerical works 
\citep[c.f.][]{Governato04,Governato07,Scannapieco06,Scannapieco08,
CK2009,Zavala08,DallaVecchia08,Stinson09}. 
We have experimented by switching off cooling 
\citep[c.f.][]{GI97,TC00,Stinson06} 
by 40 Myr in order to allow for proper
expansion of the heated regions. This time is roughly the time
a pressure--driven super-shell will attain $\gsim 150$ pc size
and the time a star of eight solar masses live.
When this feedback
is allowed in our low--mass galaxy, the stellar effective radius increases by a factor of 
$\sim 2$, the galaxy becomes less concentrated, and its $V_c(R)$ profile 
becomes flatter than in the weak-feedback simulation (c.f. runs D2 
and D3). {\it A stronger feedback helps also to keep an episodic
SFH}. A complex multi--phase structure in the disk is favored
by the action of feedback but if it is too strong, the low--density, warm and hot
gas can be completely blown away from the disk, as it seems to be the 
case in run D4.

\begin{figure}[htb!]
\plotone{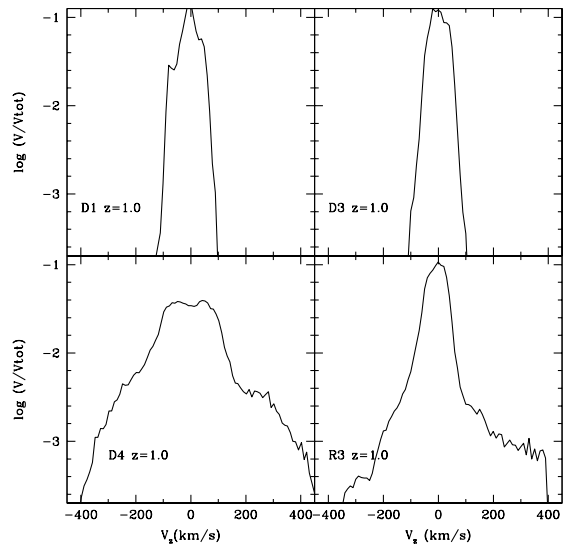}
\caption[fig:VzD]{Distribution function, weighted by volume, of the
vertical velocity component of the gas for the analyzed galaxy 
at $z= 1$ for runs D1, D3, D4, and R3. We have used all cells
within \rv. As was already seen in Fig.\ \ref{fig:VzImage} runs
D4 and R3 exhibit outflow velocities that go up to 400 $\kms$ .
\label{fig:VzD}}
\end{figure}

The warm and hot gas seen above the disk plane can be produced by two
mechanisms: (1) shocks of the infalling gas, and/or (2) gas heated and ejected from 
the mid-plane by supernovae and stellar winds. Which is the dominant process depends on
the infalling history of the gas and on the efficacy of the stellar feedback. For example,
in run D1 (weak feedback) it is very unlikely that the hot gas seen at intermediate 
altitudes ($\sim 1-3\ \kpc$) is produced by the second mechanism. 
On the other hand, the increasing amount of hot gas seen at very high altitudes 
(as high as $\gsim 5\ \kpc$) in runs D4, R3, and R4 at $z = 1$ appears to come from 
the second mechanism. This is confirmed visually and by the analysis of the 
gas $v_z$ distribution, where $v_z$ is the gas velocity component perpendicular to 
the plane of the disk. Figure \ref{fig:VzImage} shows run R3 at $z = 1$. The
projected spatial distribution of the gas temperature and 
of the $z$ velocity component are shown in left and right panels, respectively. 
Figure \ \ref{fig:VzD}, on the other hand, shows the $v_z$
normalized distribution of runs D1, D3, D4, and R3 at $z = 1$. 
Here we have used all cells inside \rv. Unlike runs D1 and D3, runs D4 and 
R3 exhibit high $|v_z|$ wings, corresponding to a galactic wind.

\subsubsection{Self--regulated SF and the multi--phase ISM}

Several of the results described above can be understood as a consequence of
the efficiency by which the thermal energy is injected into the ISM
and the ability of the medium to self--regulate its SF.
The R2--R4 random--SF sequence (cooling turned off after SF events)
shows that the reduction in $\nsf$ causes: (a) a flatter circular velocity profile; 
(b) a stronger spheroidal stellar component; (c) a thicker disk;
and when run R2 is compared with R3 or R4, (d) a delay of the
maximum of SFR.
The deterministic--SF sequence shows that, at fixed $\nsf$ (runs D2, D3, and D4),
a greater ability of SF to heat the gas causes a more efficient self--regulation of 
SF, preventing the gas from going into stars too rapidly. This gives flatter $V_c(R)$
profile and allows the gas to settle into an extended disk configuration.
However, when feedback is strong enough to produce galactic superwinds, 
the self--regulation can be interrumped and SF proceeds in a bursty regime. 
Indeed, run D4 is the one where ejecting feedback is most efficient and it is also the one 
with the most bursty SFH.

A multi--phase structure of the galactic ISM is not easy to attain in cosmological
numerical simulations \citep[see for a discussion, e.g.,][]{Stinson06,CK2009}. 
For example, we see
in run D1 that only the cold phase is present in the mid-plane of the disk while
in run D2 the hot phase is absent (see Figure \ref{fig:FFD}). These two
runs have an inefficient feedback, but even in those for which the cooling is switched off
for 40 Myr after each SF event (strong feeback, runs D3 and D4), 
the hot phase is also absent. The only run in which a  
a multi--phase medium is observed at the analyzed epochs is R2.
Its density and temperature PDFs show wide and nearly bimodal
distributions, with different gas phases, from cold--warm dense gas to
hot low--density gas, coexisting within the galaxy. The increase of the
SFR with time of run R2 is more gradual 
and not too bursty, which suggests a more uniformly self--regulated SF process
with only small episodic phases. Finally, this run takes the longest
time to assemble the disk.

In summary, the results of our experiments show that a multi--phase ISM in low--mass galaxies
is not a straightforward product of feedback and resolution. While the former
certainly helps, in low--mass systems
it also produces outflows that leave only cold/warm high--denstiy gas in the galaxy and 
induce a bursty SFH.   

Finally, in our simulations the \sigSFR--\siggas\ relation tends to be 
significantly steeper than the empirical determinations for normal disk galaxies, 
which are typically explained by the quiescent, self--regulated 
nature of the SF process \citep[e.g.,][]{Kennicutt98,HAF01,Martin01}. 
For reference, we plot in Figure \ref{fig:SigmaSFR} 
the fit to observations by \citet{Kennicutt98} for averages values of normal disk 
galaxies, their centers, and starbust galaxies (solid line). This correlation
is roughly maintained for azimuthally averaged elements along a given disk \citep{Martin01}.  
As can be seen, most of our runs at both $z=1$ and $z=0.43$ show a steeper 
local correlation 
than the observational fit of normal galaxies. Overall, most of the simulation points 
lie around the observational fit. The points from run D1, which is the run that suffers 
most from the overcooling problem, are the most distant from the fit. It is interesting 
to note that among LSB galaxies and along their azimuthally averaged elements, the 
\sigSFR--\siggas\ correlation has been found to be significantly steeper than the given by 
Kenniccutt \citep{Wyder09}. LSB galaxies are typically of low mass. 

\begin{figure*}
\includegraphics[width=\textwidth]{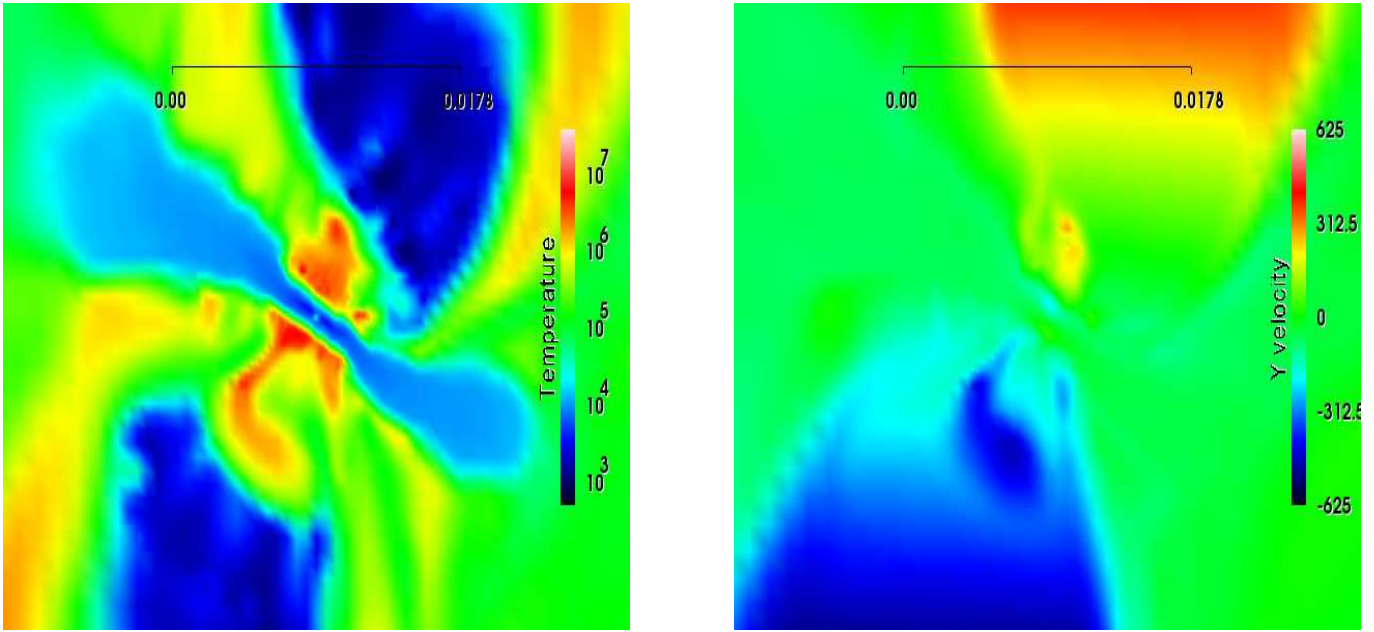}
\caption[fig:VzImageD4]{Edge-on view of model
D4 at $z = 0.43$. {\it Left panel:} Distribution 
of the gas temperature. {\it Right panel:} Distribution of the 
velocity component perpendicular to the disk. Slices are 
25 comoving kpc on a side (length of solid line) and 219 
comoving pc thick. It can be
seen that the gaseous disk is warped and more homogeneously
distributed than that shown in Figure \ref{fig:VzImage}. Gas
velocities are no longer structured in shells and velocities  
exceed 300 $\kms$ far from the disk.
\label{fig:VzImageD4}}
\end{figure*}

\subsubsection{The simulation with run--away stars (run R1)}

The subgrid physics used in run R1 differs from the rest of simulations
explored herein. In run R1 the radiative cooling is not switched off but a  
signicant fraction of stars are modeled as runaway stars able to migrate to 
lower gas density regions. The SN explosions in the lower density gas, where
the cooling time is long, are more efficient. This model has been succesful in
reproducing many of the properties of Milky Way-sized galaxies \citep[][]{CK2009, cdb2009}. 
As shown here, it works
reasonably well also for a low-mass galaxy. It successfully reproduces a
roughly flat $V_c(R)$ profile and a multiphase ISM at the
galaxy mid-plane. The model also has the lowest galaxy mass baryon fraction
and relatively high angular momentum content. On the other hand, its SFR was
relatively too efficient at early epochs (the first 1-2 Gyr). As a result, at
later times, $z < 1$, the SFR was lower than other `feedback-efficient' models,
and the model ends up with a lower fraction of mass in gas (see, for example, 
R1 versus D3). These effects are probably caused by the high value used for 
the star formation efficiency, $C_*$. Overall, run R1 exhibits similar
properties to those of run D3 and D4 as a result of its ability to
heat the gas efficiently through the runaway stars.

\subsection{Galaxy assembly in low--mass \LCDM\ halos}

Our simulations suggest that the effects of feedback
are more relevant for the galaxy morphology than in massive systems. The main
effects are {\it galaxy outflows and a trend to form a significant extended stellar 
spheroid with an approximately exponential surface density profile}. 
What is the origin of this stellar halo? Since the dark matter halo under study
evolves in a quiet fashion, at least from $z \sim 2$ to $z \sim 0.4$, without
experimenting major mergers, a significant fraction of the
stellar particles of this extended spheroid were born in situ and 
{\it not by mergers} as it is expected for large galaxies. This 
is confirmed by our analysis of tracing back in time the 
position  of the stellar particles in order to infer their origin. 
Recently, \citet{Stinson09} 
have also reported the formation 
of old extended exponential stellar halos in their simulations of dwarf--like galaxies
in isolated (without mergers) halos. As these authors review, observations show 
that dwarf galaxies are characterized by extended exponential old/intermedium 
age stellar halos. 
A recent study of LSB galaxies with the stacked image technique has also 
shown the existence of red faint stellar halos around LSB galaxies
\citep{Bergvall09}, galaxies that are supposed to form typically in isolated environments.
It is worth noticing that the R runs (those with a stochastic SF model), 
with the exception of R2, produce more massive and extended stellar halos 
than the D runs (those with deterministic SF model), as seen in Fig.\ \ref{fig:SurfDensity}.

The parameters of the SF/feedback prescriptions used here were shown to 
affect significantly the evolution and properties of the modeled low--mass galaxy.
The change of these parameters affects in a complex and non--linear way the
evolution and properties of the system. We note that even the evolution of
the dark matter mass and angular momentum distributions is different for different 
combinations of the SF/feedback parameters, confirming the 
existence of a {\it tight evolutionary 
interconnection between baryons and dark matter} 
\citep[see for recent results, e.g.,][]{Weinberg08,Abadi09,Pedrosa09}.
In particular, the internal distribution of the angular momentum evolves in our 
simulations, as can be seen, in particular, by changes in
the dark  halo and baryonic galaxy 
specific angular momentum ratio.
We suggest that our finding that the baryon specific angular momentum can 
be larger than  the one in the dark halo, deserves further future exploration.

In order to obtain a not too concentrated low--mass galaxy, with a nearly flat rotation 
curve, multi--phase ISM, and an extended and episodic SFH, an efficient stellar feedback
is necessary (in our case, switching off radiative cooling in the star forming 
region by 40 Myrs). However, the SF parameters are also important; for the resolution 
we have attained in our simulations (nominally 218 comoving pc), effective values for 
the density 
threshold $\nsf\sim 1-10 \pcc$ are recommended.
We say `effective'  because in the stochastic (random) SF scheme
the reported $\nsf$ is just the minimum value, larger 
values of $\nsf$ are actually chosen, and SF proceeds with a 
probability distribution given 
by eq. (\ref{eq:prob_SF}). 
As in previous related numerical treatments
of the galaxy problem, we conclude that sub--grid physics
needs to be carefully addressed (tuned) in order to obtain reasonably realistic galaxy 
properties.

\subsection{Shortcomings}

Despite that several
of the runs presented here predict realistic properties\footnote{The aim of 
the present work was not to produce a realistic galaxy.}, we note that the circular 
velocity curves are decreasing for all the cases studied here, while in observed 
low--mass galaxies the trend is to have an increasing {\it rotation curve} up to the 
last measured radii \citep[c.f.][ but see Swaters et al. 2009]{Persic96}. 
However, the average rotational
velocity profile does not trace necessarily the circular velocity and 
at least in model R2 
the rotational velocity, of the gas at $T < 10^4$ K,  increases 
as observed in low-mass galaxies. 
Similar situations have been discussed by \citet{Valenzuela07} 
as a result of pressure gradients and non-circular motions.
The mismatch between circular and rotation velocity deserves
further exploration in future papers exploring also effects of
resolution.

Also, our simulated galaxies show baryonic
mass fractions that are too high compared with observational inferences. 
For example, in the simulation where
galaxy outflows are most effective (run D4), $f_{gal}(5\re) = 0.06$ (at $z=0.43$) is only  
a factor 2.5 lower than the universal baryon fraction $f_b$, 
while the observational inferences suggest that present--day galaxies of low masses 
have baryon fractions more than a factor of 5 smaller than $f_b$. 
Our results confirm that {\it producing  a stellar feedback 
that is sufficiently efficient as to 
eject considerable amounts of gas from even low--mass galaxies is not an 
easy task} \citep[see also][ and references therein]{DT2008}.

Even more, the problem of producing strong enough outflows is probably
in opposition with other potential vexing conflict. An increasing amount of 
local and high--redshift observational evidence shows that the less massive the
galaxies, the higher their current specific SFRs (SSFR = SFR/$M_*$) on average at 
all redshifts back to $z \sim 1-2$, a phenomenon dubbed as `downsizing in 
SSFR' \citep{Fontanot09}. Moreover, the observational inferences show that the 
problem is not that massive disk galaxies have low SSFRs but that 
low--mass galaxies have too high SSFRs 
\citep[see for a discussion][ and references therein]{Firmani09}.
Such SSFRs imply that low--mass galaxies should delay the formation of stars and
do it with a nearly constant or even increasing SFRs \citep{Noeske07}.
On the other hand, if $f_{gal}$ decreases as the galaxy mass decreases, then 
obtaining late, high SSFRs in these galaxies becomes even more difficult. 
As concluded in \citet[][ see also Somerville et al. 2008 and Fontanot et al. 2009]{Firmani09} 
both problems are difficult to solve in the context of \LCDM--based models. 

The cosmological numerical simulations presented here confirm the sharpness
of both potential problems. {\it The baryon fractions of the simulated low--mass galaxies
are too high and their SSFRs are in all cases much smaller than the average values 
of the observational inferences at redshifts $\sim 0.4-1$}. The run with the highest
SSFRs is D3:  SSFR = 0.17, 0.06 Gyr$^{-1}$ at $z=1$, 0.43, 
respectively.
Observations of large samples of late--type galaxies at high redshifts report 
values $\sim 5$ times higher for the corresponding $z'$s and stellar masses 
\citep[][ see also Fig. 1 in Firmani et al. 2009]{Bell07, Zheng07, Noeske07, Damen09}.
Most  of the other runs have SSFRs much smaller than those
of run D3. Our simulations also show that if $f_{gal}$ is reduced by strong
feedback--driven outflows, then the late-time SSFR decreases.  
Run D4 indeed has fewer baryons owing to the strong outflows, but also a much 
lower SSFR than run D3. A related problem to the low SSFRs is that {\it the gas fractions 
of the simulated galaxies (for example at $z = 0.43$) are in most cases 
too low} (see Table \ref{tab:BFract}). Among runs R2--R4, the maximum value of $f_g$ is 0.075 
for run R2; among runs D2-- D4, as expected by its SSFR history, run D3 has the 
largest value, $f_g=0.37$, which is still marginal compared with the large gas 
fractions of observed present--day low--mass galaxies. They should have large 
gas reservoirs to produce the high SSFRs infered from observations.

 If the baryon fraction and SSFR problems for low--mass galaxies are confirmed,
models for SF/feedback very different to the one applied here ---and in other works---
should be proposed to agree with observations, preserving
the underlying \LCDM\ cosmogony.  It is possible that significant changes
to the galaxy assembly process can be introduced alternatively by improving the 
physics of the poorly understood circumgalactic and intergalactic media. Finally,
we should say that the current observational inferences, specially at high
redshifts, are still plagued by large uncertainties \citep[][]{Conroy09a,Conroy09b,
Firmani09}. 

A comparison of 
simulation outcomes at $z=0$ with the more accurate observations of present--day 
galaxies is desirable in order to asses the severity of the problems mentioned above. 
We plan to do this in a forthcoming paper.  

Finally, it should be mentioned that the sub-grid prescriptions in our
simulations are certainly not complete, as several physical ingredients
that could be important for the evolution of real galaxies are
missing. In particular, our simulations do not include the magnetic
field nor cosmic ray pressure, and they do not resolve the whole
plethora of phenomena that occur at the scales of molecular clouds,
including turbulence and other forms of stellar feedback additional to
supernova explosions, such as bipolar outflows and the expansion of HII
around massive stars. All of these agents contribute to drive the
turbulence at scales that are unresolved in our simulations, whose
effect may be to reduce the star formation efficiency in the simulations
\citep{KHM00, VS_etal03, MK04, RG_etal10, VS_etal10}. The magnetic field
is expected to have a similar effect \citep{OGS99, HMK01,
VS_etal05}. Note that, in a sense, this is what we are accomplishing in
our simulations by turning off the local cooling after a star formation
event, since the hotter gas is less prone to forming stars. However,
future studies will need to quantify these effects so that they can be
incorporated in a suitable sub-grid model applicable for
galaxy-formation simulations.

\section{Conclusions}

By means of high--resolution AMR hydrodynamic simulations (the hydrodynamics 
+ N-body ART code), 
we have explored the formation and evolution of a galaxy within a low mass 
halo ($\sim 7 \times 10^{10}\msunh$ at $z=0$) assembled in the context of the 
\LCDM\ cosmogony. The DM particle mass of the least massive species
is $5.3 \times\ 10^5\ \msunh$.
Our main aim was to explore the formation of a low--mass galaxy 
in cosmological simulations with {\it the same initial conditions} but 
varying the most relevant parameters of 
the SF and feedback recipes. The analysis presented herein was mainly made
at two epochs, $z=1.0$ and 0.43, a time period of 3.2 Gyr for the used cosmology. 
From this exploratory study --not exaustive in any sense-- we remark the following
systematic results:
  
- The SF density threshold $\nsf$ is a relevant parameter in the simulations, 
which affects the main structural and dynamical properties of the stellar and gaseous components
of the low--mass galaxies, as well as their SFHs and general ISM properties. As $\nsf$ is
decreased, (a) the stellar galaxy size ($R_e$) increases; (b) the $V_c(R)$ profile becomes 
less peaked; (c) the peak in the SFH is delayed and the late (nearly exponential) 
decline in SFR becomes more gradual; (d) the stellar populations are younger on average, with a thicker
disk and a larger fraction of stars in a ``puffier'', spheroidal component; (e) the 
low--density and high--temperature gas in the disk increases. For a cell of 150 pc, 
close to our nominal resolution at $z=0$, we estimate from typical ISM properties that 
$\nsf$ should be $\sim 5\ \pcc$.

- The stellar particle mass limit, $\mlim$, affects the efficiency of SN energy
dumping into the cell. For high values of $\mlim$, the feedback--driven disk outflows
increase and the SFH of the simulated galaxy becomes burstier. 

- An effective stellar feedback (generated by turning off locally the cooling by 40 Myr 
after SF) has a strong effect on our studied low-mass galaxy: the stellar radius $R_e$ increases by
a factor of $\sim 2$, the galaxy becomes less concentrated, and $V_c(R)$ becomes flatter.
A strong feedback produces also an episodic SFH. Its action also favors the development
of a complex multi--phase disk ISM and a self--regulated SF process, but if it is too 
strong, the low--density, warm/hot gas can be completely blown out from the disk.  

- For all of our runs, both at $z=1$ and 0.43, we find that the relation between 
\sigSFR\ and \siggas\ is steeper than empirical determinations for normal disk galaxies, but
in agreement with recent determinations for LSB galaxies, which are typically of low mass.
Nevertheless the effective stellar surface densities, $\Sigma_M$, of our models are 
instead in the high--density side of the observed distributions at 
$z\approx 0.4$ and 1.

- In all runs, the stellar galaxy has a significant extended spheroidal component 
(between $\sim 30$ and $60\%$ of $M_*$ at $z=0.43$) with a nearly exponential surface density
profile. We have seen that an important fraction of the stars in this halo were formed in situ,
while a minor fraction may come from minor merger events. 

- The variation of the parameters of the SF/feedback prescriptions affects in a complex
and non--linear way the evolution of the galaxy system. Even the evolution of the dark mass
and angular momentum distributions change as these parameters are changed, confirming the 
tight interconnection between baryons and dark matter. For example, the internal distribution 
of AM evolves in our simulations and this evolution is sensitive to the SF/feedback physics.

- In all runs of the low--mass galaxy studied here, the baryon fractions are
higher, the gas fractions are lower, and the SSFRs are much smaller than the corresponding 
average values inferred from observations. In particular, the
too low SSFRs values found in our model galaxies pose a sharp problem to the 
simulations and the model in general
if the yet uncertain observational inferences are confirmed.

\section*{Acknowledgments}

We are grateful to A. Kravtsov for providing us with the numerical code and for
help with its usage. 
We are in debt to N. Gnedin for providing us the analysis and graphics 
package IFRIT.
The authors acknowldege PAPIIT-UNAM grants IN112806 to P.C., IN114509 to V.A., 
and IN118108 to O.V., CONACyT grants 60354 (to P.C., V.A., and
O.V) and U47366-F (to E.V.-S.), and the Lady Davis Trust to D.C. for partial funding. 
We also acknowldege the anonymous referee for a careful reading of the manuscript.
Some of the simulations presented in this paper were performed on the HP CP 
4000 cluster (Kan-Balam) at DGSCA-UNAM.

\label{lastpage}

\end{document}